# Pattern Formation in *E. coli* Through Negative Chemotaxis: Instability, Condensation, and Merging

Nir Livne, Ady Vaknin, and Oded Agam

The Racah Institute of Physics, Edmond J. Safra Campus, The Hebrew University of Jerusalem, Jerusalem 9190401, Israel.

**Abstract**

Motile bacteria can migrate along chemical gradients in a process known as chemotaxis. When exposed to uniform environmental stress, *Escherichia coli* cells coordinate their chemotactic responses to form millimeter-sized condensates containing hundreds of thousands of motile cells. In this study, we combined experiments with mathematical modeling based on modified Keller-Segel equations to investigate the dynamics of this collective behavior across three distinct time scales: the shortest time scale, where spatial instability emerges; an intermediate time scale, where quasi-stationary bacterial condensates form; and finally, a longer time scale, during which neighboring bacterial accumulations coalesce. The model closely agrees with experimental results, quantitatively capturing the observed instability, the shape of mature condensates, and their coalescence dynamics. Specifically, we found that the force between neighboring bacterial accumulations decays exponentially with distance due to screening effects. We suggest that the model presented here could describe more broadly the dynamics of stress-induced condensation mediated by bacterial chemotaxis.

## I. INTRODUCTION

Motile bacteria navigate their environments by sensing and responding to chemical gradients in a process termed chemotaxis. This response involves moving toward beneficial substances (attractants) and away from harmful ones (repellents) [1-3]. Chemotaxis can also give rise to collective behaviors when cells modify their chemical environment in ways that lead to coordinated responses. Collective behaviors driven by chemotaxis along attractant gradients have been previously demonstrated and theoretically analyzed [4-12]. Pattern formation mediated by negative chemotaxis was recently demonstrated when *Escherichia coli* (*E. coli*) cells were immersed in an initially uniform acidic environment, which acts as a repellent field [13]. In this set up, the neutralization of repellent molecules by the cells creates an attraction between them. When exposed to an acidic environment, *E. coli* adopts various strategies to regulate its internal pH [14-17]. The mechanism pertinent to this behavior involves the absorption of protons from the surrounding environment and their neutralization, leading to an increase in the external pH. In regions where bacterial density is spatially heterogeneous, this local neutralization induces pH gradients that neighboring cells can sense. The combination of diffusion, local pH regulation, and the bacteria's tendency to migrate toward less hostile environments destabilizes the uniform bacterial distribution, resulting in the formation of distinct bacterial accumulations separated by characteristic distances [13]. Over longer periods, neighboring bacterial accumulations that are sufficiently close tend to coalesce. This condensation of bacteria into dense communities reduces the stress they endure and can lead to the formation of tightly packed aggregates resembling biofilm structures. Therefore, condensation likely plays an important ecological role, contributing to the survival of cells under various stress conditions. However, a comprehensive quantitative understanding of these phenomena is still lacking.

*E. coli* is a micron-sized bacterium equipped with several helical flagella that drive it forward when spinning counterclockwise, in a movement called a "run" [18-20]. When some flagella spin clockwise, the bacterium reorients its direction in a process called a "tumble". Alternating sequences of runs and tumbles, occurring without any external cues, result in a random walk. During a run, *E. coli* moves in relatively straight lines at a typical velocity of $v \sim 25\,\mu\text{m/sec}$ for a period about $\tau_{run} \sim 1\,\text{sec}$, whereas tumbles last only a fraction of that time, typically around one-tenth of a second [21, 22]. Thus, bacteria can be associated with an active diffusion constant of about $D = v^2 \tau_{run}$.

*E. coli* senses the environment through specialized chemoreceptors that detect chemical signals and trigger a cascade of intracellular events to adjust the bacterium's movement [23]. The binding of attractants increases the likelihood of runs, while the binding of repellents causes more frequent tumbling events [24]. To cope with varying concentrations of chemicals, dedicated proteins adapt the chemoreceptors, enabling the bacteria to detect and respond to relative changes in concentration (log-sensing) over a wide dynamic range [25-27]. Since the adaptation process is much slower than the immediate responses to chemical signals, there is a lag between the two effects, serving as a form of "memory" that allows cells to measure spatial gradients along their run length. By constantly sampling its surroundings, *E. coli* extends runs when conditions improve and increases tumbles when they worsen, effectively biasing its random walk toward more favorable environments.

The Keller-Segel equations describe the dynamics of chemotactic agents in a chemical environment using reaction-diffusion equations [28]. These equations are often used to model attractant-mediated condensation [11, 29]. We use a modified version of this model, adapted to bacteria confined to move within a domain $\mathcal{D}$ immersed in a three-dimensional gel with an initially uniform low pH level [13]:

$$\frac{\partial c}{\partial t} = D_c \nabla^2 c - \alpha \rho c \chi_{\mathcal{D}}(\boldsymbol{R}), \tag{1.1a}$$

$$\frac{\partial \rho}{\partial t} + \nabla \cdot \boldsymbol{J} = 0, \text{ with } \quad \boldsymbol{J} = -D\nabla\rho - \kappa\rho\nabla\ln(K+c), \text{ and } \quad \boldsymbol{R} \in \mathcal{D}. \tag{1.1b}$$

In these equations, $c$ and $\rho$ represent the densities of protons and bacteria, respectively. The first equation (1.1a) describes the dynamics of the protons, which includes diffusion with a diffusion constant $D_c$, and accounts for removal of protons by the bacteria, effectively acting as sinks. These sinks are present only within the domain $\mathcal{D}$, where the bacteria are confined—a constraint enforced by the characteristic function $\chi_{\mathcal{D}}(\boldsymbol{R})$, where $\boldsymbol{R}$ is the three-dimensional spatial coordinate. This function equals one if $\boldsymbol{R} \in \mathcal{D}$, and zero otherwise. In our experiments, $\mathcal{D}$ is a thin, quasi-two-dimensional layer. Finally, the parameter $\alpha$ governs the rate at which protons are neutralized by the bacteria. In the proton density regime relevant to our system, $\alpha \propto \sqrt{c}$ exhibits only a weak dependence on $c$ [13]. For simplicity, we neglect this weak nonlinearity and treat $\alpha$ as a constant. This approximation has little effect on the results (see Appendix G for further discussion).

The second equation, (1.1b), is the continuity equation for the bacteria. This equation is applicable when bacterial growth is arrested, and the bacteria do not die during the experiment's duration. The bacterial flux, $\boldsymbol{J}$, includes two components. The first is due to the gradient of bacterial concentration, resulting in diffusive dynamics with an active diffusion constant $D$. The second component represents chemotactic drift, meaning the bacterial flux is directed along the gradient toward regions of lower repellent concentration, where the parameter $\kappa$ controls the effectiveness of chemotaxis. Since bacteria respond to relative changes of protons over a wide dynamic range, their drift velocity can be modeled as being proportional to $\ln(K+c)-\ln(K'+c)$, where $K$ and $K'$ are the lower and upper cutoffs, respectively [30-32]. For $c \ll K'$ the second term can be neglected, resulting in the logarithm in Eq. (1.1b). Further details on the system setup and the equations of motion are provided in Appendix A.

In the following study, we provide a quantitative understanding of stress-induced bacterial condensation using dedicated experimental tools combined with an analytical analysis of Eqs. (1.1). Specifically, we first determine the conditions for the onset of condensation. Next, we calculate the morphology of a mature bacterial condensate, and finally, we investigate the interactions between distinct bacterial condensates, focusing on their coalescence dynamics. Our theoretical findings, which are in quantitative agreement with the experimental data, improve our understanding of the dynamics within the unstable bacterial state. The insights gained from this study are valuable for predicting bacterial behavior under different stress conditions, with potential applications not only in advancing our understanding of bacterial ecology but also in practical areas like managing bacterial contamination of medical devices and in the food industry.

## II. RESULTS

In this section, we present our experimental findings along with their corresponding theoretical interpretations based on the analysis of Eqs. (1.1). These results encompass the chemotactic instability observed at short time scales, the formation of distinct cell condensates at intermediate time scales, and their eventual coalescence over longer time scales.

### II.1 The chemotactic instability

We begin by demonstrating the chemotactic instability of bacteria under an initially uniform stress. Following Livne et al. [13], a drop of liquid containing freely swimming *E. coli* bacteria was placed onto acidic agarose hydrogel in conditions that do not support cell proliferation (Fig. 1a). Within 20 minutes, the bacteria condensed into isolated accumulations, with a typical distance between neighboring condensates of approximately 1.7 mm (Fig. 1b-c). The power spectrum of the resulting pattern, shown in Fig. 1c, displays a single ring, indicating the lack of long-range order within the pattern. These millimeter-sized accumulations, composed of hundreds of thousands of motile cells, remain dynamic and interact with neighboring condensates, resulting in their eventual coalescence (Fig. 1d).

To determine the conditions required for the onset of the instability and to identify its characteristic wavelength, we perform a linear stability analysis of Eqs. (1.1). Notably, unlike condensation due to secretion of attractant molecules, the steady-state solution to Eqs. (1.1) is uniform, with the repellent depleted and cells dispersed. However, if protons are present in the initial conditions, instability may transiently arise and persist for hours [13]. We assume the bacteria occupy a quasi-two-dimensional region with a uniform density $\rho_0$, located on top of an acidic hydrogel. This hydrogel is assumed to fill the entire lower half-space, with an initial spatially uniform proton density $c_0 \gg K$. For this geometry and initial conditions, Eqs. (1.1) do not admit a uniform fixed-point solution. Therefore, the chemotactic instability arises on top of a time-dependent but spatially uniform solution:

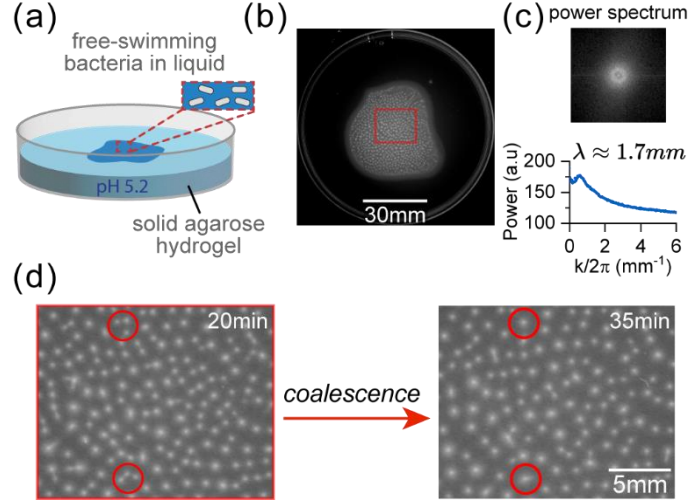

**Figure 1. Pattern formation in the density of *E. Coli* due to negative chemotaxis.** By consuming repellent molecules, bacteria attract each other to form dense accumulations [13]. **(a)** An illustration of the experimental setup. Chemotactic *E. coli* cells were suspended in an acidic motility buffer with low buffering capacity (pH 5.2, 0.1mM KPi) and spread over agarose hydrogel made of the same motility medium. The applied suspension was thin (~500 μm). See Methods. **(b)** Top view of the bacterial suspension at $t = 20$ min displaying the formed pattern. **(c)** Top: Power spectrum of the pattern shown in (b). Bottom: radial profile of the power spectrum. $\lambda$ corresponds to the wavelength at the peak. **(d)** Magnified view of the pattern inside the rectangle marked in (b) is shown on the left along with a similar image of that region at $t = 35$ min on the right. Red circles display groups of distinct condensates that have coalesced into one. Evidently, a pattern of cell condensates is formed, and these distinct accumulations later coalesce as cohesive groups.

$$\rho = \rho_0, \text{ and } c_{\mathcal{D}}(t) = c_0 \exp\left(\frac{t}{\tau}\right) \operatorname{erfc}\left(\sqrt{\frac{t}{\tau}}\right). \tag{2.1}$$

Here, $c_{\mathcal{D}}(t)$ is the proton density within the quasi-two-dimensional region occupied by the bacteria, whose thickness, $h$, is assumed to be much smaller than all other relevant length scales in the problem. The time scale characterizing the changes in the average proton concentration at the bacterial layer is given by:

$$\tau = \frac{D_c}{\alpha^2 h^2 \rho_0^2}. \tag{2.2}$$

The solution above is derived under the assumption that the thickness of the agarose hydrogel beneath the cells is infinite. If the thickness, $W$, is finite, Eq. (2.1) remains valid for $t < \tau_W$, where $\tau_W = W^2/D_c$ represents the typical time required for protons to diffuse across a distance equal to the thickness of the cell-free hydrogel. For $t \gg \tau_W$ the decay of $c_{\mathcal{D}}(t)$ becomes exponential, and any accumulated cells will eventually disperse as $c$ drops below $K$. In our setup, $\tau_W$ is estimated to be approximately one day, which is significantly longer than the duration of the experiment. Further details regarding the derivation of Eq. (2.1) are provided in Appendix B.

Given that the instability of the uniform solution (2.1) develops on a timescale much shorter than $\tau_W$, a linear stability analysis of this solution with respect to non-uniform spatial perturbations reveals that the most unstable mode corresponds to the wavenumber

$$k_{ins} = \frac{2\pi}{l_{ins}} = \frac{\sqrt{\alpha \kappa \rho_0}}{2D}, \qquad (2.4)$$

which defines the characteristic length scale of the spatial instability, $l_{ins}$. In this analysis, we assume $D_c \simeq D$ in accordance with the experimental situation [13, 33]. A generalization of the instability analysis for other ratios of $D_c/D$ (see Appendix C), shows that the system exhibits slightly increased instability when $D_c \neq D$.

Eq. (2.4) can be interpreted as follows: Consider a uniform bacterial density perturbed by a small spatial modulation with amplitude $\delta \rho$ and spatial period $l_{ins}$. The typical time for the proton distribution to become spatially modulated due to this perturbation is $1/(\alpha \delta \rho)$. This time should be comparable to the time it takes for protons to diffuse over a distance comparable to the spatial period of the modulation, which is $l_{ins}^2/D$. Setting these times equal gives: $\delta \rho \sim D/(\alpha l_{ins}^2)$. On the other hand, the relative perturbation $\delta \rho / \rho_0$ arises from the competition between chemotaxis, which tends to condense the cells (governed by the parameter $\kappa$), and diffusion, which works to disperse them (characterized by $D$). Since both $\kappa$ and $D$ have the same physical units, the balance between these processes yields $\delta \rho / \rho_0 \sim \kappa/D$. By equating the two expressions for $\delta \rho$, we obtain $l_{ins} \sim D/\sqrt{\rho_0 \alpha \kappa}$ which up to numerical constant of order unity matches the length given by Eq. (2.4).

We conclude this subsection with two comments. First, in systems with a lateral dimension $L$, boundary conditions impose a lower limit on the possible values of $l_{ins}$ - instability only occurs when $l_{ins} < L$. In other words, chemotactic instability is suppressed if the lateral system size is too small, as observed in experiments [13]. Second, the above instability analysis is valid as long as the characteristic time $\tau$ for changes in the proton concentration at the bacterial layer is much longer than the typical time required for the instability to develop, which is given by $\tau_{ins} = 4D/(\alpha \kappa \rho_0)$. Namely, the stability analysis holds when $\tau/\tau_{ins} = \kappa/(4h^2 \alpha \rho_0) \gg 1$, a condition always satisfied if the cell layer is thin enough. Further details of the linear stability analysis are provided in Appendix C.

## II.2 The density profile of a single bacterial condensate

The linear stability analysis described above outlines the system's dynamics within a short period, just a few minutes after the bacteria are placed on the acidic agarose hydrogel. It demonstrates that a spatial instability with a characteristic length scale $l_{ins}$ develops and grows exponentially over a time scale of approximately $\tau_{ins}$. Over a significantly longer period, this instability evolves into distinct bacterial accumulations, which eventually coalesce over even larger time scales (as detailed in the next subsection). Our goal here is to characterize the density profile of a single bacterial accumulation.

We conduct condensation experiments in a thin well imprinted into agarose hydrogel [13]. This setup allows us to resolve the condensation dynamics at high spatial and temporal resolutions. Fig. 2a presents a fluorescence image of bacteria from a condensation experiment in a channel geometry. The time evolution of the bacterial density is characterized in Fig. 2b, which displays the density profile as a function of position along the channel at three time points. Initially, perturbations to the cell density formed, with peaks spaced approximately 2-2.5 mm apart, as predicted by linear stability analysis ($l_{ins}$

= 2.2 ± 0.7 mm). These initial perturbations continued to grow while maintaining their inherent spacing, and after approximately one hour, the density profile stabilized and became nearly time independent.

The shape of mature condensates can be predicted by theory. Fig. 2c displays the bacterial density of these mature condensates, normalized by setting $\rho_N(x) = [\rho(x) - \rho_\infty]/(\rho_{max} - \rho_\infty)$ where $\rho_\infty$ represents the residual density far from the accumulation, $\rho_{max}$ is the maximal density, and $x$ is the distance from the condensate center. The colored dots display the experimentally measured density along the central line of the channel for six mature condensates from three different biological samples, while the red line represents the approximate theoretical profile in a quasi-steady state:

$$\rho_N(x) \simeq \left(1 + \frac{x^2}{\ell^2}\right)^{-\xi}, \text{ with } \xi = \frac{\kappa}{D}, \qquad (2.5)$$

and

$$\ell = \sqrt{\frac{N\xi}{\pi h \rho_{max}}}. \qquad (2.6)$$

Here, $N$ is the number of bacteria within a condensate. Details of the analytical derivation of the above formulas are provided in Appendix D, with numerical support in Appendix E. In Fig. 2, $\ell$ and $\xi$ are used as fitting parameters, with the best fit obtained for $\ell = 0.55 \text{mm}$, and $\xi = 6$. Note that the spatial coordinates of all data points are not rescaled.

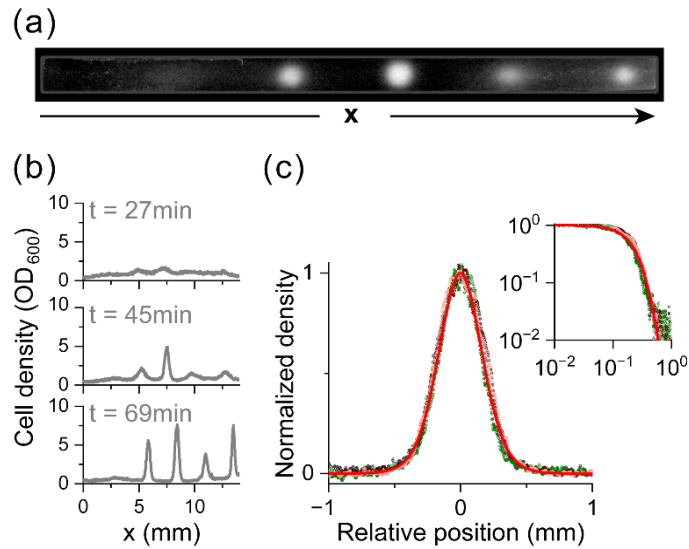

**Figure 2. The density profile of mature cell condensates in a channel geometry.** Cells at an initial $OD_{600}$ of 0.8 were suspended in a low buffering capacity acid motility medium and restricted to move inside rectangular wells ($14 \times 0.8 \times 0.1\,mm^3$). The wells were formed in agarose hydrogel made of the same medium as the cells (see Methods). **(a)** Top view fluorescence image of a typical cell layer at $t = 60$ min, shown in log-scale for clarity. The long side of the rectangle is defined as the $x$-axis. **(b)** Cell density profiles along the x-axis at various times (as labeled) of the experiment shown in (a). The fluorescence intensity was measured along center of the well. **(c)** The shape of bacterial condensates. The density profiles of two mature condensates from three distinct biological replicates (n = 6) are plotted in different colors. Position was shifted relative to the center of condensates, and their profiles were normalized by subtracting the baseline cell density and dividing by the peak density (see text). The analytic approximation $\rho_N = (1 + x^2/\ell^2)^{-\xi}$ of the shape is shown in red ($\ell = 0.55$ mm, $\xi = 6$; see text). Inset – log-log plot of data and theory. Overall, condensates matured within about 60 min, with a shape predictable by theory.

Assuming the number of bacteria within a condensate is effectively determined by the instability length scale, $l_{ins}$, we find that in a stripe geometry, this number is given by $N = l_{ins} w h \rho_0$, where $w$ is the channel width, taken to be much smaller than $l_{ins}$. Substituting this estimate into Eq. (2.6), we obtain the theoretically predicted condensate size $\ell \approx l_{ins}/3 = 0.6 \pm 0.2$ mm, which is consistent with the experimental observations.

Despite the good fit presented in Fig. 2c, when the total cell count within a condensate exceeded a certain threshold ($N \approx 2 \times 10^6$ cells), deviations emerged between the shape of the bacterial accumulation and the theoretical profile. These deviations were observed both in a larger, disk-like (2D) bacterial layer (Figs. 8-9 in Appendix D) and in a channel geometry with higher initial cell density (Fig. 10 in Appendix D). In both scenarios, condensates broadened, with their shapes diverging from the theoretical prediction. Given that the numerical solution of Eqs. (1.1) aligns with the theoretical approximation (Fig. 11 in Appendix E), this divergence at high cell counts suggests the influence of additional factors not accounted for in the model.

**II.3 Coalescence of distinct condensates**

Cells form condensates through self-generated modulation in the distribution of protons. The same mechanism can, in principle, also create attractive forces between separate cell accumulations. These interactions may cause neighboring condensates to move toward one another, eventually leading to their merger, as demonstrated in Ref. [13] and Fig. 1d.

To study these coalescence dynamics, we devised a dedicated experimental setup in which pairs of condensates are formed at a controlled distance between them. Bacterial condensation occurs regardless of the shape of the well in which cells are suspended, but their spatial arrangement can be controlled by modifying the lateral boundaries. As depicted in Fig. 3a, we triggered the formation of two bacterial accumulations at desired positions using an agarose well consisting of two triangles connected by a thin channel. This setup promoted a dynamic evolution in two stages: initially, two nearly identical accumulations formed, and subsequently, the attraction between them led to their coalescence (Fig. 3a-c, Fig. 13 in Appendix F, and SI Video 1).

Using the density profiles of interacting condensates we measured the distance between their centers of mass, denoted hereinafter by $2a$, as a function of time (Fig. 3c). Viewing the dynamics of distinct bacterial accumulations as those of two overdamped (identical) particles, the relative velocity of their movement is proportional to the attractive force between them. Therefore, to measure the force acting between condensates, we calculated the velocity $2da/dt$. In Fig. 3d, we plot the results as a function of the distance between the centers of mass of the accumulations. These data indicate that the attractive force between merging condensates decays exponentially with distance, suggesting that in the dynamics, screening effects play a dominant role. Interestingly, even when their tails were already overlapping, distinct condensates maintained their shape while moving (Fig.13 in Appendix F).

The coalescence of condensates is driven by their interaction rather than the geometry of the experimental setup. This conclusion is supported by observations in a setup containing a single triangular well, where a condensate formed but did not exhibit directed movement (Fig. 12a in Appendix F). Similarly, in a configuration where two triangular wells were connected by a longer (1 mm) channel, condensates formed independently within each well but remained approximately 3 mm apart without attracting each other (Fig. 12b in Appendix F). These findings indicate that the attraction between condensates arises from their mutual interaction, which, within the duration of our experiments, is effective over a range of approximately 2 mm.

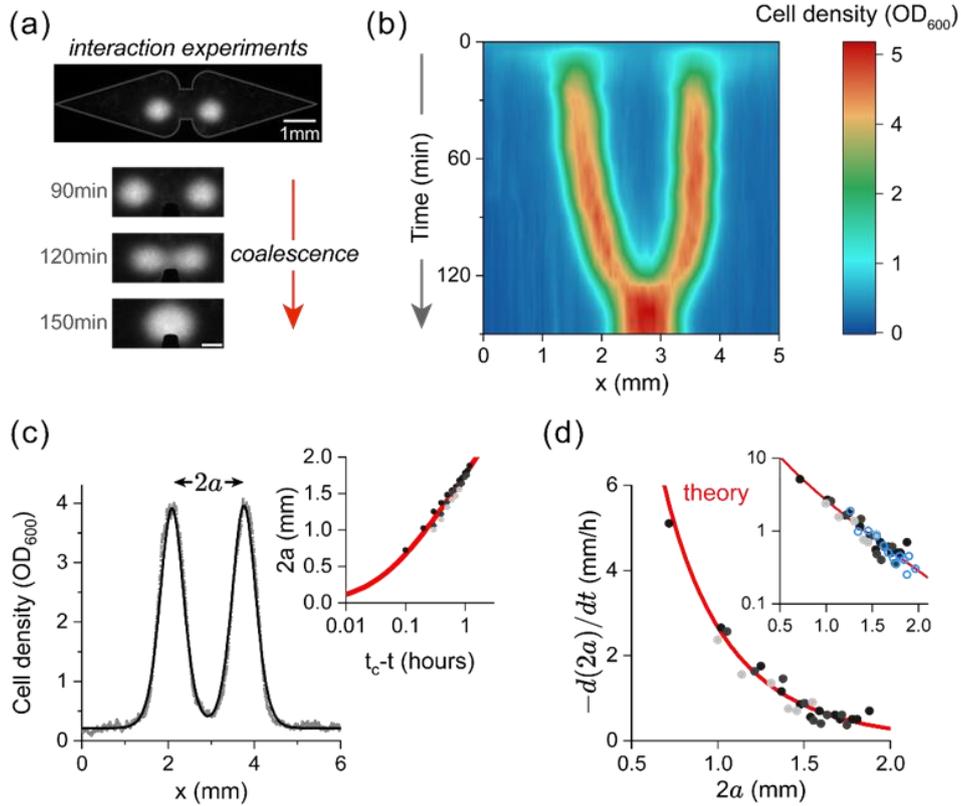

**Figure 3. Coalescence dynamics of distinct bacterial accumulations.** Cells were suspended in an acidic motility medium at an initial OD$_{600}$ of 0.6 and constrained to swim within a setup made of two triangular shapes connected by a narrow channel imprinted in agarose hydrogel. This setup induces bacterial condensation in controlled positions and allows their physical contact. **(a)** Snapshot image of the bacterial layer at $t$ = 90min. Two mature condensates can be seen, and the time course of their coalescence is demonstrated below (times as labeled; scale bar: 0.5mm). **(b)** Kymograph of the cell density along the axis connecting the two condensates. **(c)** Quantitative analysis of the dynamics of coalescing condensates. The cross-sectional cell distribution at t = 90min is plotted along with the theoretical approximation of their shape (see Fig. 2 and text). Inset - Distance ($2a$) between the accumulations plotted against the time until coalescence ($t_c$, time of coalescence). Gray markers display the experimental data from three biological replicates and the red line is a theoretical fit to $2a(t) \simeq \ell \ln\left[\left(\tau_{\text{merge}} + t_c - t\right)/\tau_{\text{merge}}\right]$ with $\ell = 0.55$ mm (as in Fig. 2) and $\tau_{\text{merge}} = \ell^2/4D = 0.042$ hours. **(d)** The force between coalescing condensates. The velocity $2da/dt$ is plotted as a function of distance $2a$. Three biological replicates of the experiment are shown as shaded gray markers. The red line shows the theoretical fit to the data $-\ell K_1(2a/l)/\tau_{\text{merge}}$. Inset – semi-log plot of experimental data and the theoretical fit. Data from experiments with the two disconnected triangles is shown as open blue markers (see text and SI Video 2). Evidently, the force between mature cell condensates decays exponentially with the distance between them.

To further examine this interaction, we repeated the merger experiments using a setup similar to that in Fig. 3a but with a hydrogel barrier placed between the two triangular wells. Despite this separation, condensates continued to attract each other (blue markers in the inset of Fig. 3d; see also SI Video 2). Because the hydrogel barrier is permeable to protons but not to bacteria, hydrodynamic interactions could be ruled out. Therefore, we conclude that the observed attraction between distinct condensates is driven by chemotaxis via self-generated proton gradients.

Using the chemotaxis model described above, we can calculate the time dependence of the distance between the condensates, $2a(t)$. The derivation assumes a quasi-stationary state and that each bacterial condensate is localized with a well-defined center of mass. The resulting equation for the distance (see Appendix F) is:

$$\frac{d2a}{dt} = -\frac{\ell}{\tau_{merge}} K_1\left(\frac{2a}{\ell}\right) \tag{2.7a}$$

with

$$\tau_{merge} \simeq \frac{\ell^2}{4D} \tag{2.7b}$$

up to a constant of order one. Here $K_1(z)$ is the modified Bessel function of the second kind. The exponential decay of the function $K_1(z)$ for large values of $z$ is the manifestation of the aforementioned screening effects, with screening length approximately equal to the size of the condensate, $\ell$. The time scale $\tau_{merge}$ is approximately the duration required for information about the proton field to diffuse from the edge of a condensate to its center. By measuring the value of $\ell$ from the size of the condensates found earlier (Fig. 2), and using the estimated value of $D$ ([33]; Table 1 in Appendix F), the two parameters in formula (2.7b) are determined with no other fitting parameters, and the resulting function is shown as the solid line in Fig. 3d. Integration of Eq. (2.7a) (see Appendix F) yields the approximate formula:

$$2a(t) \simeq \ell \ln\left[\frac{1}{\tau_{merge}}\left(\tau_{merge} + t_c - t\right)\right], \tag{2.8}$$

where $t_c$ is the merging time. This function is depicted by the solid line in the inset of Fig. 3c, together with the experimental data points.

**III. DISCUSSION**

Bacteria can respond to uniform stress by forming dense accumulations through feedback between their ability to deplete repellent molecules and their migration down repellent gradients. Consequently, a thin, uniformly distributed bacterial layer in contact with a large reservoir of acidic pH becomes unstable (Figs. 1-2). This instability results in the formation of isolated cell accumulations (Figs. 1-2), which create favorable conditions with reduced stress for cells in each condensate [13]. The characteristic length and timescales of this instability arise from a balance between chemotaxis, which promotes bacterial accumulation, and diffusion, which disperses the bacteria. The effective attraction between cells, mediated by the proton field, also generates an attractive force between entire bacterial accumulations (Fig. 3). However, this attraction between distinct condensates decays exponentially with distance, limiting condensate-merging events. This screening effect accounts for the observed spacing between condensates and the lack of long-range order in their arrangement.

Bacterial condensation occurs only if cells collectively respond to their self-induced gradients before the overall repellent concentration falls below detectable levels. When cells occupy only a small portion of the environment, repellent removal occurs on a longer time scale than bacterial condensation, enabling the development of instability from an initially uniform distribution. In a finite system, however, the repellent concentration will eventually drop below the threshold required for effective chemotaxis, depending on the dynamic range for the corresponding sensory response [13], and would eventually lead to dispersal of the accumulated cells (Fig. 7 in Appendix C). Additionally, a finite bacterial population can only modify its surroundings to a limited extent, and once saturation is reached, diffusion smooths out any inhomogeneities. Thus, although negative chemotaxis instability may persist for several hours, it remains an inherently transient phenomenon.

The theoretical model used here captures the emergence of a quasi-stationary state following the onset of instability and provides formulas for both the density profile of individual bacterial accumulations and the attractive force between neighboring condensates. This model can be readily adapted to different bacterial species and stress conditions. By predicting the conditions under which bacterial condensates form and coalesce, the model offers practical applications, such as assisting in the development of strategies to inhibit condensation-triggered aggregation and its potential development into biofilm on medical devices and food surfaces.

Despite its success in capturing key aspects of condensate dynamics, our model has several limitations and challenges, as it omits additional biological factors that may influence population behavior. In particular, we neglect effects that become significant at high cell densities, such as aerotaxis and volume-filling constraints, both of which can broaden the bacterial distribution. These effects are especially relevant when the total cell count is high, either due to the initial conditions (see Figs. 8–10 in Appendix D) or as a result of condensate coalescence (see Fig. 3 and Ref. [13]). Consequently, our model is only valid for describing the dynamics before these extreme densities are reached. Additionally, the model does not account for the secretion of other biochemical effectors, such as the quorum-sensing molecule autoinducer-2, which could influence both the spatial distribution and temporal evolution of bacterial condensates [34, 35]. Incorporating these biological factors in future refinements would enhance the model's predictive power, particularly in more complex environments.

In summary, we investigated the dynamics of bacterial condensation under uniform acidic stress using experiments and mathematical modeling. The model quantitatively predicts the onset of instability, the shape of condensates at the quasi-steady state, and the coalescence of condensates. The agreement between theory and experiments suggest that the model can be generalized to other stress conditions and may have practical applications in understanding bacterial organization and colonization in such environments.

## IV. MATERIALS AND METHODS

**Strains and Plasmids.** We used the wild-type *E. coli* MG1655 (VF6) expressing GFP from pTrc99a vector induced with 75 μM IPTG (pSA11).

**Cell growth and motility medium**. Cells were grown over night in Tryptone Broth (TB, tryptone extract 10 g/l and 5 g/l NaCl) supplemented with Amp, diluted 100-fold into fresh medium with IPTG (75 μM), and allowed to grow aerobically at 33.5°C with shaking. Cells were harvested at an $OD_{600}$ of 0.5 by centrifugation, gently washed twice in motility medium, and resuspended in motility medium. The motility medium we used included: 0.1 mM potassium phosphate (KPi), 0.94 mM lactic acid, 85.5mM sodium chloride, 0.1 mM EDTA, and 1 μM methionine, titrated to pH 5.2 with NaOH. Notably, this medium supports long-term bacterial motility but not cell proliferation.

**Agarose plate assay.** Standard 90 mm plates were filled with agarose hydrogel (1%, in motility medium, pH 5.2). Once the agarose solidified, the gel-filled plates were dialyzed overnight in motility buffer and a second time, for an additional 4 hours, in a fresh motility buffer just before the experiment (final dialysis ratio of $10^2$:1). GFP-expressing bacteria at $OD_{600}$ 1 suspended in motility buffer were then applied by pipetting 1 ml of the bacterial suspension on top the hydrogel and allowed to expand and form a thin layer. Plates were kept covered and at room temperature (24˚C). To observe the fluorescence distribution, plates were illuminated using uniform LED illumination at 470 nm (runVIEW Mini Blue, Cleaver Scientific), and viewed through a dedicated filter screen using a DLSR camera (Canon EOS2000D). The images in Fig. 1 present the green channel of the RGB image.

**The 'thin layer' setup**. Agarose gel (3%, in motility medium, pH 5.2) was casted in a cylindrical chamber made from grade-2 titanium (height: 6 mm) and sealed at the top using a mold with a thin (~ 100 μm) protrusion. These molds, with protrusions of various shapes, were fabricated using a resin printer (Elegoo Mars 3). After the agarose solidified, the mold and the bottom cover were removed and the agarose was dialyzed overnight in motility buffer and a second time, for an additional 4 hours, in a fresh motility buffer just before the experiment (final dialysis ratio of $10^4$:1). Bacterial cells were then introduced into the shallow well in the agarose hydrogel formed by the molds, and the bottom and top of the chamber were covered with BSA-treated glass coverslips (30min treatment in 1% BSA) and secured in position. The setup was kept at room temperature (24˚C) for observation. The redistribution of the cells was followed by tracking changes in the fluorescence distribution under the microscope.

**Fluorescence microscopy**. The bacterial distribution was followed using a Nikon Ti microscope equipped with a 4× (0.2 NA) objective and a decoded controlled stage. Large images were obtained by stitching smaller images (approximately 2mm x 2mm each) with 20% overlap using a built-in function in NIS Elements software (Nikon). The bacterial distribution inside the well was extracted from the images by removing the measured background intensity and normalizing to the average intensity within each well. Radial averaging (data Figs. 7, 8 and 10) was done by measuring the fluorescence intensity around the center of cell accumulations using Radial Profile Angle plugin for ImageJ.

**Acknowledgments**

We thank the Milner Foundation for partly funding NL's fellowship. This work was supported by the Israeli Science Foundation and the Minerva Center for Bio-Hybrid Complex Systems.

**Author contributions**

NL and AV designed the experiments. NL performed the experiments and numerical calculations. OA performed the analytical calculations. NL, AV and OA analyzed the data and wrote the manuscript.

**Competing interests**

The authors declare no competing interests.

## V. APPENDICES

### APPENDIX A: THE SYSTEM SETUP & EQUATIONS OF MOTION

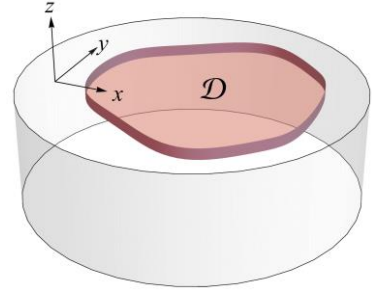

**Figure 4.** An illustration of the experimental setup.

An illustration of the experimental setup is shown in Fig. 4. In this setup, bacteria are restricted to a narrowly defined domain $\mathcal{D}$, whereas protons are free to diffuse across the entire lower half-space, $z<0$, which includes the region occupied by bacteria. In what follows, $\rho$ and $c$ will denote the densities of bacteria and protons (in units of inverse volume), respectively.

The proton's density satisfies the equation

$$\frac{\partial c}{\partial t}=D_c\nabla^2 c-\rho r_c\chi_\mathcal{D}(\boldsymbol{R}); \quad \chi_\mathcal{D}(\boldsymbol{R})=\begin{cases}1 & \boldsymbol{R}\in\mathcal{D}\\ 0 & \boldsymbol{R}\notin\mathcal{D}\end{cases}. \tag{A1}$$

Here $\boldsymbol{R}$ denotes a three-dimensional position coordinate, $D_c$ represents the diffusion constant of the protons, For the rate of repellent uptake, we use the approximation $r_c=\alpha c$ which can be considered as a low-concentration limit of a Michaelis-Menten term in which $\alpha$ is a parameter that characterizes the rate at which bacteria remove protons them from the system. The function $\chi_\mathcal{D}(\boldsymbol{R})$ is the characteristic function, which equals one if $\boldsymbol{R}$ resides with the domain inhibited by the bacteria and zero otherwise. The condition that there is no proton flux flowing into the upper half space sets the boundary condition:

$$\left.\frac{\partial c}{\partial z}\right|_{z=0}=0. \tag{A2}$$

It is assumed that the small thickness of the domain $\mathcal{D}$, henceforth denoted by $h$, ensures that diffusion renders the bacteria and the proton densities, within the domain $\mathcal{D}$, essentially independent of the perpendicular coordinate, $z$.

Choosing $\boldsymbol{r}$ to be a coordinate vector in the reduced space occupied by the bacteria, the equations governing the bacteria dynamics is

$$\frac{\partial \rho(\boldsymbol{r},t)}{\partial t}=\nabla\cdot\{D\nabla\rho(\boldsymbol{r},t)-\rho(\boldsymbol{r},t)\boldsymbol{v}_d\}=0. \tag{A3}$$

where $D$ is the diffusion constant of the bacteria, and

$$\boldsymbol{v}_d=-\kappa\nabla\ln\left[\frac{K+c_\mathcal{D}(\boldsymbol{r},t)}{K'+c_\mathcal{D}(\boldsymbol{r},t)}\right] \tag{A4}$$

is the drift velocity due to chemotaxis, in which $\kappa$ characterizes the strength of chemotaxis, and

$$c_\mathcal{D}(\boldsymbol{r},t)=c(\boldsymbol{R},t)\big|_{z=0}. \tag{A5}$$

In the above expression for the drift velocity, $K$ and $K'$ characterize the sensory response of the cells to a given the repellent (here protons). For typical repellents, $K'\gg K$ and thus the bacteria are essentially log-sensing within a wide dynamic range. Moreover, since repellents can have deleterious effects on the bacteria at high concentrations, we assume that, as in all our experiments, $c \ll K'$, and approximate the drift velocity as

$$\boldsymbol{v}_d \simeq -\kappa \, \nabla \ln\left[K + c_{\mathcal{D}}(\boldsymbol{r},t)\right] . \tag{A6}$$

The parameter $K$ represents a small constant density that acts as a regularizer which helps to avoid divergent solutions in the distribution of bacterial colonies.

The initial conditions are of uniform proton distribution in the lower half space, and uniform bacteria density within the domain $\mathcal{D}$:

$$c(\boldsymbol{R},t)\big|_{t=0} = c_0; \ z<0, \quad \text{and} \quad \rho(\boldsymbol{r},t)\big|_{t=0} = \rho_0 . \tag{A7}$$

Finally, in our analysis, we set $D_c = D$ to align with the conditions observed in the experiments.

**APPENDIX B: SOLUTION FOR SPATIALLY UNIFORM BACTERIAL DENSITY**

In this appendix, we construct solutions for $c(\boldsymbol{R},t)$ under the condition that the bacteria density remains uniform in space, $\rho(\boldsymbol{r},t) = \rho_0$ (for any $t$). We shall consider the two cases depicted in Fig. 5. In the first case, the bacteria are uniformly distributed across a two-dimensional plane, while the second case considers bacteria confined within a quasi-one-dimensional domain. In both cases, it is assumed that the lateral extent of the system is infinite. Given this assumption, the symmetry of the problem permits satisfying the boundary condition (A2) by symmetrically extending the system to the upper half space, as illustrated in Fig. 5.

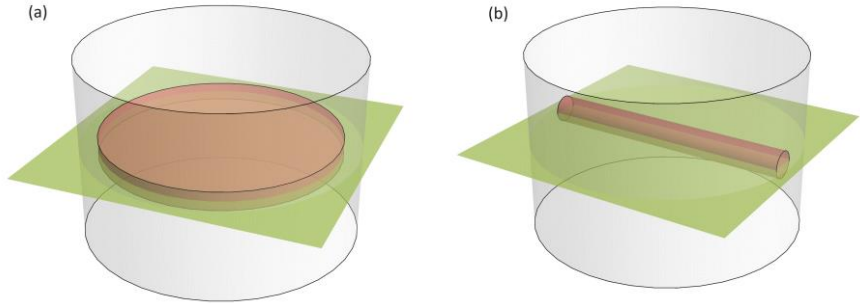

**Figure 5.** Symmetrization of the system into the upper half space for the case where bacteria occupy a quasi-two-dimensional space (a) and a quasi-one-dimensional space (b).

To begin, consider the case where bacteria are uniformly distributed within a quasi-two-dimensional plane of thickness $h$. Let $\bar{c}$ represent the solution for the proton density in this scenario. To simplify the analysis, the characteristic function $\chi_{\mathcal{D}}(\boldsymbol{R})$ is approximated by $h\delta(z)$, thereby reducing Eq. (A1) to:

$$\frac{\partial \bar{c}}{\partial t} = D\frac{\partial^2 \bar{c}}{\partial z^2} - \alpha 2h\rho_0 \bar{c} \delta(z) . \tag{B1}$$

In this equation, the factor of 2 in the second term on the right-hand side accounts for the symmetrization of the model into the upper half space.

For this geometry, it is advantageous to employ dimensionless variables $c' = \bar{c}/c_0$, $t' = t/\tau^{2D}$, and $z' = z/\sqrt{D\tau^{2D}}$, with

$$\tau^{2D} = \frac{D}{\alpha^2 h^2 \rho_0^2}. \tag{B2}$$

By taking the Laplace transform of the corresponding dimensionless equation with respect to time, we obtain the equation

$$\left(s - \frac{\partial^2}{\partial z'^2} + 2\delta(z')\right)\hat{c}(z';s) = c'(z',0) = 1, \tag{B3}$$

where

$$\hat{c}(z';s) = \int_0^\infty dt' c'(z',t')\exp(-st'). \tag{B4}$$

To solve Eq. (B3) we define the Green function of the problem, $\mathcal{G}(z,z_0;s)$, which satisfies the equation:

$$\left(s - \frac{\partial^2}{\partial z'^2} + 2\delta(z')\right)\mathcal{G}(z',z_0';s) = \delta(z' - z_0'), \tag{B5}$$

The Dyson equation for this Green's function is

$$\mathcal{G}(z',z_0';s) = \mathcal{G}_0(z',z_0';s) - 2\int d\tilde{z}\,\mathcal{G}_0(z',\tilde{z};s)\delta(\tilde{z})\mathcal{G}(\tilde{z},z_0';s), \tag{B6}$$

where $\mathcal{G}_0(z',z_0';s)$ is the free Green's function that satisfies the equation:

$$\left(s - \frac{\partial^2}{\partial z'^2}\right)\mathcal{G}_0(z',z_0';s) = \delta(z' - z_0'). \tag{B7}$$

From Eqs. (B6), it follows that

$$\mathcal{G}(0,z_0';s) = \mathcal{G}_0(0,z_0';s) - 2\mathcal{G}_0(0,0;s)\mathcal{G}(0,z_0';s). \tag{B8}$$

Solving for $\mathcal{G}(0,z_0';s)$ and substituting the result back to Eq. (B6), we obtain:

$$\mathcal{G}(z',z_0';s) = \mathcal{G}_0(z',z_0';s) - 2\frac{\mathcal{G}_0(z',0;s)\mathcal{G}_0(0,z_0';s)}{1 + 2\mathcal{G}_0(0,0;s)}. \tag{B9}$$

With the help of this Green's function, the solution of Eq. (B3) is given by

$$\hat{c}(z';s) = \int_{-\infty}^\infty dz_0' \left[\mathcal{G}_0(z',z_0';s) - 2\frac{\mathcal{G}_0(z',0;s)\mathcal{G}_0(0,z_0';s)}{1 + 2\mathcal{G}_0(0,0;s)}\right]. \tag{B10}$$

Substituting

$$\mathcal{G}_0(z',z_0';s) = \int_0^\infty dt' \frac{1}{\sqrt{4\pi t'}}\exp\left(-\frac{(z'-z_0')^2}{4t'} - st'\right) = \frac{\exp(-\sqrt{s}|z'-z_0'|)}{2\sqrt{s}}, \tag{B11}$$

and performing the integral over $z'_0$ yields the Laplace transform of the proton distribution:

$$\hat{c}(z';s) = \frac{1}{s}\left[1 - \frac{\exp(-\sqrt{s}|z'|)}{1+\sqrt{s}}\right]. \tag{B12}$$

In particular, for $z' = 0$ (the domain inhibited by the bacteria)

$$\hat{c}(0,s) = \frac{1}{s+\sqrt{s}}, \tag{B13}$$

and the inverse Laplace transform yields the proton density within the domain $\mathcal{D}$ given in Eq. (2.1). This solution has the following asymptotic behaviour:

$$c_{\mathcal{D}}(t) \simeq c_0 \begin{cases} 1 - 2\sqrt{\dfrac{t}{\pi \tau^{2D}}} & t \ll \tau^{2D} \\ \dfrac{1}{\sqrt{\pi t/\tau^{2D}}} & t \gg \tau^{2D} \end{cases} \tag{B14}$$

Thus for $t \gg \tau^{2D}$, the proton density within the domain occupied by the bacteria becomes a slowly varying function. This property is essentially the reason for the quasi-stationary behavior observed in the system during the time interval $\tau^{2D} \ll t \ll \tau_W$, where $\tau_W$ represents the time required for protons to diffuse across the system, assuming it has a finite size $W$.

The function $c'(z',t')$, evaluated numerically from the inverse Laplace transform of $\hat{c}(z;s)$, is depicted in Fig. 6. Panel (a) illustrates the proton density as a function of the dimensionless distance $z' = z/\sqrt{D\tau^{2D}}$ for various time points, evenly spaced between $t' = 0$ and $t' = 4$. Here, early times are represented by blue curves, while red colors indicate later times. Panel (b) displays the time dependence of the proton density at $z' = 0$.

The solution of Eq. (1.1) for the case of the channel geometry depicted in Fig. 2a, with a uniform bacterial distribution, is challenging. To simplify the problem. we consider the case where the bacteria occupy a quasi-one-dimensional domain, which, in the symmetrized system, is modeled as a cylinder with radius $h$ as illustrated in Fig. 5b. Assuming the cylinder extends along the $x$ direction, we redefine the coordinates such that $\boldsymbol{R} = (x, \boldsymbol{u})$, with $\boldsymbol{u} = (y, z)$. In this setup, the characteristic function $\chi_{\mathcal{D}}(\boldsymbol{R})$ is replaced by $\pi h^2 \delta(\boldsymbol{u})$, and the equation for the proton density reduces to:

$$\frac{\partial \bar{c}}{\partial t} = D\frac{1}{u}\frac{\partial}{\partial u}\left(u\frac{\partial \bar{c}}{\partial u}\right) - \alpha\pi h^2 \rho_0 \bar{c}\,\delta(\boldsymbol{u}). \tag{B15}$$

Using the rescaled variables $\boldsymbol{u}' = \boldsymbol{u}/\sqrt{D}$ and $c' = c/c_0$, and following the same steps as for the previous case, we obtain:

$$\hat{c}(\boldsymbol{u}';s) = \int d^2u'_0 \left[\mathcal{G}_0(\boldsymbol{u}',\boldsymbol{u}'_0;s) - \eta\frac{\mathcal{G}_0(\boldsymbol{u}',0;s)\mathcal{G}_0(0,\boldsymbol{u}'_0;s)}{1+\eta\mathcal{G}_0(0,0;s)}\right], \tag{B16}$$

where $\eta = \pi h^2 \alpha \rho_0 / D$ is a dimensionless parameter, and

$$\mathscr{G}_0(\mathbf{u}', \mathbf{u}'_0; s) = \int_0^\infty dt \, \frac{1}{4\pi t} \exp\left(-\frac{(\mathbf{u}' - \mathbf{u}'_0)^2}{4t} - st\right) = \frac{1}{2\pi} K_0\left(|\mathbf{u}' - \mathbf{u}'_0|\sqrt{s}\right) \quad (B17)$$

is the Laplace transform of the diffusion Green's function in two-dimensional space. Here, $K_0(x)$ is the modified Bessel function of second kind and zeroth order. This function diverges logarithmically near the origin, therefore, $\mathscr{G}_0(0,0;s)$ should be regularized by setting the rescaled minimal distance to be $u'_{min} = h/\sqrt{D}$. With this regularization:

$$\hat{c}(u;s) = \frac{c_0}{s} \left[1 - \frac{\frac{\eta}{2\pi} K_0\left(u\sqrt{\frac{s}{D}}\right)}{1 + \frac{\eta}{2\pi} K_0\left(h\sqrt{\frac{s}{D}}\right)}\right]; \, u \geq h. \quad (B18)$$

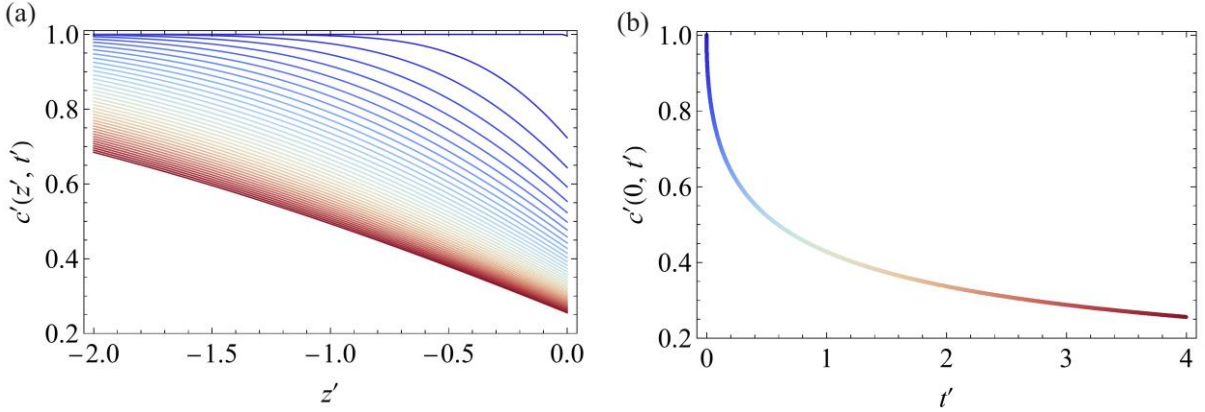

**Figure 6. The proton density distribution for a time-independent uniform bacterial distribution.** The dimensionless proton density, calculated from the inverse Laplace transform of (B12), as a function of the distance, $z$, from the plane occupied by the bacteria. (a) Each curve in the panel represents a different time point with a thermometer color code in which blue is associated with early time and red with later time. (b) The dimensionless proton density at point $z=0$ as a function of the dimensionless time. The color code is the same as in panel (a).

Evaluating this function on the cylinder, $u = h$ in the asymptotic limit $s \to 0$, which corresponds to long-time behavior, yields

$$\hat{c}(h;s) \xrightarrow[s \to 0]{} \frac{c_0}{s} \frac{2\pi}{2\pi - \frac{\eta}{2} \ln\left(\frac{h^2 s}{4D \exp(-2\gamma)}\right)}, \quad (B19)$$

where $\gamma = 0.5772...$ is Euler's constant. Taking the inverse Laplace transform of this expression yields the asymptotic behavior in the long-time limit:

$$c_{\mathcal{D}}(t) = \bar{c}(h,t) \xrightarrow{Dt \gg h^2} \frac{1}{2\pi} \lim_{Y \to \infty} \int_{-Y}^{Y} \frac{ds}{is+\varepsilon} \frac{2\pi c_0}{2\pi - \frac{\eta}{2}\ln\left(\frac{h^2(is+\varepsilon)}{4D\exp(-2\gamma)}\right)} \exp(ist+\varepsilon t) \quad (B20)$$

$$\simeq \frac{c_0}{1 + \frac{\eta}{4\pi}\ln\left(\frac{t}{\tau^{1D}}\right)},$$

where

$$\tau^{1D} = \frac{h^2}{4\exp(2\gamma)D}. \quad (B21)$$

This result indicates that in the quasi-one-dimensional case, the proton density remains practically constant over time when $t \gg \tau^{1D}$, i.e., for times longer than the diffusion time across the thickness of the bacterial domain.

**APPENDIX C: LINEAR STABILITY ANALYSIS**

In this appendix, we present the linear stability analysis of the uniform (time-dependent) solution derived in the previous appendix. For simplicity, in this analysis, the regularizer $K$ is set to zero. This assumption is justified given that the instability develops during the early stages of the system's dynamics when the proton concentration within the bacterial domain remains high, and the impact of the regularizer is negligible.

To be concrete, we consider the instability in the case of a two-dimensional bacterial distribution. It should be noted, however, that our findings are equally relevant to the quasi-one-dimensional geometry. We will linearize the equations of motion (A1-A3) by setting $\rho = \rho_0 + \delta\rho(\mathbf{r},t)$ and $c = \bar{c}(z,t)[1 + \delta c(\mathbf{r},t)/c_0]$. With these substitutions, the equation (A1) for the proton distribution reduces to:

$$\frac{c(z,t)}{c_0}\frac{\partial}{\partial t}\delta c(\mathbf{r},t) = \frac{c(z,t)}{c_0}D_c\nabla_\perp^2 \delta c(\mathbf{r},t) - \alpha c(z,t)\delta\rho(\mathbf{r},t)h\delta(z), \quad (C1)$$

where we approximate $\chi_{\mathcal{D}}(\mathbf{R})$ by $h\delta(z)$, $D_c$ is the protons diffusion constant, while $\nabla_\perp^2$ denotes the Laplacian in the $xy$ plane. Upon integrating this equation over the thickness of the bacterial domain, $h$, we obtain:

$$\frac{\partial \delta c^{2D}(\mathbf{r},t)}{\partial t} = D\nabla_\perp^2 \delta c^{2D}(\mathbf{r},t) - \alpha c_0 \delta\rho^{2D}(\mathbf{r},t), \quad (C2)$$

where $\delta c^{2D}(\mathbf{r},t) = h\delta c(\mathbf{r},t)$ and $\delta\rho^{2D}(\mathbf{r},t) = h\delta\rho(\mathbf{r},t)$. Similarly, Eq. (A3) for the bacteria density becomes

$$\frac{\partial \delta\rho^{2D}(\mathbf{r},t)}{\partial t} = D\nabla_\perp^2 \delta\rho^{2D}(\mathbf{r},t) + \frac{\kappa\rho_0}{c_0}\nabla_\perp^2 \delta c^{2D}(\mathbf{r},t). \quad (C3)$$

Taking the Fourier transform of Equations (C2) and (C3) with respect to $r$, we arrive at:

$$\frac{\partial}{\partial t}\begin{pmatrix} \delta\hat{c}^{2D} \\ \delta\hat{\rho}^{2D} \end{pmatrix} = \begin{pmatrix} -D_c k^2 & -\alpha c_0 \\ -k^2 \frac{\kappa\rho_0}{c_0} & -Dk^2 \end{pmatrix}\begin{pmatrix} \delta\hat{c}^{2D} \\ \delta\hat{\rho}^{2D} \end{pmatrix}, \tag{C4}$$

where $\delta\hat{\rho}^{2D}$ and $\delta\hat{c}^{2D}$ denote the Fourier transforms of $\delta\rho^{2D}$ and $\delta\rho^{2D}$, respectively. Diagonalization of the above matrix yields the eigenvalues:

$$\Lambda_\pm = k\left(-\bar{D}k \pm \sqrt{\alpha\kappa\rho_0 + \frac{\Delta D^2 k^2}{4}}\right), \tag{C5}$$

where $\bar{D} = (D_c + D)/2$, while $\Delta D = D_c - D$. The eigenvalue, $\Lambda_-$, is negative for any $k$, but $\Lambda_+$ is positive for sufficiently low values of the wave number, $0 < k < k_c$, where $k_c = \sqrt{\alpha\kappa\rho_0}/\sqrt{D_c D}$. A positive value of $\Lambda_+$ implies that the uniform solution is unstable. The maximal value of $\Lambda_+$ which governs the instability development is achieved at

$$k_{max} = \frac{\sqrt{\alpha\kappa\rho_0}}{\sqrt{2}\bar{D}}\frac{1}{\sqrt{1-\varepsilon+\sqrt{1-\varepsilon}}} = \frac{\sqrt{\alpha\kappa\rho_0}}{2\bar{D}}\left[1 + \frac{3}{8}\varepsilon + \frac{31}{128}\varepsilon^2 + \cdots\right], \text{ with } \varepsilon = \frac{\Delta D^2}{4\bar{D}^2}. \tag{C6}$$

This formula simplifies to Eq. (2.4) in the limit $q = 0$, i.e., when $D_c = D$. The corresponding maximal eigenvalue,

$$\Lambda_{max} = \frac{\alpha\kappa\rho_0}{2\bar{D}}\frac{1}{1+\sqrt{1-\varepsilon}} = \frac{\alpha\kappa\rho_0}{4\bar{D}}\left[1 + \frac{\varepsilon}{4} + \frac{\varepsilon^2}{8} + \cdots\right], \tag{C7}$$

shows that instability increases as the difference with the diffusion constants increases.

From (C6) it follows that the typical length and time scales of the instability development (in the limit $D = D_c$) are, respectively:

$$l_{ins} \sim \frac{2\pi}{k_{max}} = \frac{4\pi D}{\sqrt{\alpha\kappa\rho_0}} \quad \text{and} \quad \tau_{ins} = \frac{1}{\Lambda_{max}} = \frac{4D}{\alpha\kappa\rho_0}. \tag{C8}$$

For the experiments in Fig. 1 we find $l_{ins} = 2.0 \pm 0.7$ mm and for the experiments in Fig. 2 we find $l_{ins} = 2.2 \pm 0.7$ mm (see parameters in Table 1).

In concluding this section, it is worth noting an important point. In a finite system, chemotactic instability is inherently transient. For this instability to arise, bacteria must generate and respond to their own gradients before the overall concentration drops below detectable levels. When repellent molecules occupy a small finite volume—comparable to that occupied by the bacteria—their concentration changes quickly, which limits environmental stress and maintains a uniform distribution. However, when the repellent molecules occupy a larger volume, removal occurs over longer timescales, allowing chemotactic instability to develop. Ultimately, in a finite system, the repellent concentration will eventually fall below the threshold for effective chemotaxis, causing cells to disperse back to a uniform state, as shown in Fig. 7. This effect is not considered in the previous analysis, as we assume the repellent molecules occupy the entire lower half-space. In the experimental setup, the agarose hydrogel domain occupies a very large volume, ensuring that the negative chemotaxis instability can persist for many hours.

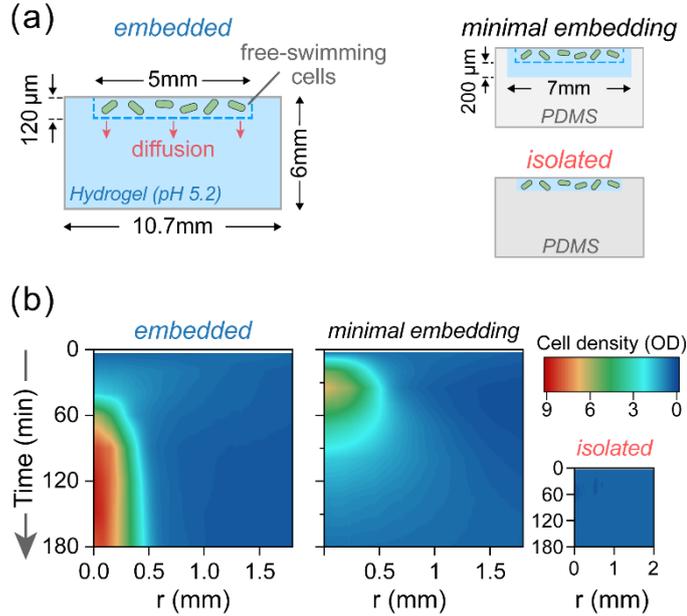

**Figure 7. The transient nature of the chemotactic instability.** (a) Schematic description of the experiments. Thin disk-like layers of GFP-expressing cells ($OD_{600} = 0.4$) were embedded within agarose hydrogel (3%) bulk of various thicknesses. Protons could diffuse freely between the hydrogel and the bacterial layer in the 'embedded' and 'minimal embedding' setups, while cells remained confined within their thin layers. Initially, a uniform acidic environment (0.1 mM KPi, pH 5.2) was established throughout the chamber, including the bacterial suspension and hydrogel. Experiments were performed with either a thick agarose hydrogel bulk ('embedded'), a thin hydrogel bulk surrounded by PDMS ('minimal embedding'), or wells entirely made of PDMS ('isolated'). Notably, PDMS allows the diffusion of oxygen gas but not protons. (b) Kymographs of cell density over time for each setup. In the embedded setup, cells condensed into a dense condensate that maintained its structure for several hours. In contrast, cells in the minimal embedding setup only formed a transient accumulation, dispersing back to a homogeneous state after about 2 hours. Isolated populations remained evenly distributed throughout the experiment.

**APPENDIX D: THE BACTERIAL CONDENSATE SHAPE**

To calculate the density profile of the bacterial condensate, we assume that time, $t$, is significantly longer than the characteristic time for proton relaxation, allowing the system to reach a quasi-steady state. Consequently, we set $\partial \rho / \partial t \simeq 0$ in Eq. (A3), with the approximation (A6), resulting in:

$$\nabla \cdot \left\{ \nabla \rho + \xi \rho \nabla \ln \left[ K + c_{\mathcal{D}}(\mathbf{r},t) \right] \right\} = 0 \qquad \text{where} \qquad \xi = \frac{\kappa}{D}. \tag{D1}$$

The expression within the curly brackets is a constant, independent of $r$. However, since both bacterial and proton densities reach constant values far from the condensate center, this constant must be zero. Thus, we obtain the equation:

$$\nabla \ln \rho(r,t) = -\xi \nabla \ln\left[K + c_{\mathcal{D}}(r,t)\right], \tag{D2}$$

whose solution is

$$\rho(r,t) = \rho_{\max}\left(\frac{K + c_{\mathcal{D}}(0,t)}{K + c_{\mathcal{D}}(r,t)}\right)^{\xi}, \tag{D3}$$

where $\rho_{\max}$ represents the maximal bacteria density achieved at the center of the accumulation, $r = 0$.

This solution should be supplemented by the condition that sets the number of bacteria within a condensate (before condensate merging occurs)

$$N = h\int d^2r\, \rho(r) \tag{D4}$$

Assuming that most of the bacteria move into the condensate, this number is determined by the instability length and the initial bacterial density, i.e., for a two-dimensional geometry

$$N = \rho_0 l_{ins}^2 h = \frac{4\pi^2 hD}{\alpha\xi} \tag{D5a}$$

while for a stripe geometry

$$N = \rho_0 l_{ins} hw = \sqrt{\frac{D\rho_0}{\alpha}}\frac{2\pi hw}{\sqrt{\xi}} \tag{D5b}$$

where $w$ is the width of the stripe which is assumed to be much smaller than $l_{ins}$. If $w > l_{ins}$ than one should use Eq. (D5a).

We now turn to consider the quasi-steady-state solution of Eq. (A1). According to Eq. (B15), the time derivative of the proton concentration $\partial c_{\mathcal{D}}(t)/\partial t$, diminishes rapidly over time, scaling as $t^{-3/2}$. This rapid decay implies that the time derivative can be disregarded when calculating the quasi-steady-state profile of the bacterial condensate. Thus assuming $\partial c/\partial t \simeq 0$ and integrating the equation along the $z$-axis across the bacterial domain yields:

$$D\frac{\partial c}{\partial z}\Big|_{-h}^{0} + hD\nabla_{\perp}^2 c_{\mathcal{D}}(r) - \alpha h c_{\mathcal{D}}(r,t)\rho(r,t) = 0. \tag{D6}$$

On the upper surface of the system, i.e. $z = 0$, we have $\partial c/\partial z = 0$ because there is no proton flux crossing that boundary. On the other hand, the proton current flowing into the bacterial domain from the proton reservoir located beneath it,

$$J_{in} = -D\frac{\partial c}{\partial z}\Big|_{z=-h}, \tag{D7}$$

is finite. Thus Eq. (D6) simplifies to:

$$\frac{J_{in}}{h} + D\frac{1}{r}\frac{d}{dr}\left(r\frac{dc_{\mathcal{D}}(r)}{dr}\right) - \alpha c_{\mathcal{D}}(r)\rho_{\max}\left(\frac{K + c_{\mathcal{D}}(0)}{K + c_{\mathcal{D}}(r)}\right)^{\xi} = 0, \tag{D8}$$

where we have substituted Eq. (D3) for the bacteria density and assumed radial dependence of the proton density (which also approximately characterizes the case of channel geometry).

This equation can be approximately solved using the following functional form, which enables matching the proton density behavior both near the condensate center and far from it:

$$c_D(r) \simeq K\left[a + (b-a)\tanh^2\left(\sqrt{\frac{\alpha\rho_{max}}{D}}qr\right)\right], \tag{D9}$$

where $a$, $b$ and $q$ are dimensionless parameters that need to be determined using Eq. (D8). The first equation for these parameters is derived by substituting the approximate solution (D9) in Eq. (D8) and taking the limit, $r \to \infty$:

$$\nu \equiv \frac{J_{in}^\infty}{J_K} = b\left(\frac{1+a}{1+b}\right)^\xi, \quad \text{where} \quad J_K = h\alpha\rho_{max}K, \tag{D10a}$$

and $J_{in}^\infty$ represents the proton flux far from the center of the bacterial accumulation.

An additional equation is obtained by substituting Equation (D9) into (D8), and demanding that it is satisfied in the limit $r \to 0$. This condition leads to the next equation:

$$4(b-a)q^2 = a - \nu, \tag{D10b}$$

Finally, we impose the constraint (D4), by substituting (D9) in (D4) and integrate over space. To simplify the calculation, we take into account that the condensate size is smaller than $l_{ins}$ (as well as channel width, $w$, in the case of a stripe geometry), and that most of the bacteria are concentrated within the condensate. This allows us to approximate the integral (D4) by expanding the hyperbolic tangent function to leading order in $r$ and extending the integration domain to include the whole two-dimensional space. This procedure leads to the equation:

$$N = h\int_0^\infty 2\pi r dr \rho(\mathbf{r},t) = \rho_{max}\left(\frac{1+a}{1+a+(b-a)\frac{\alpha\rho_{max}}{D}q^2r^2}\right)^\xi = \frac{\pi(1+a)hD}{(b-a)(\xi-1)q^2\alpha} \tag{D10c}$$

Solving Eqs. (D10) for $a$, $b$ and $q$ yields the expression for $c_D(r)$ which is then substituted into (D3) to obtain the bacterial distribution function:

$$\rho = \rho_{max}\left[1 + \frac{b-a}{1+a}\tanh^2\left(\sqrt{\frac{\alpha\rho_{max}}{D}}qr\right)\right]^{-\xi}. \tag{D11}$$

Here, we aim to solve Equations (D.10) under the condition where $\nu \to 0$, indicating that the incoming flux, $J_{in}^\infty$, is negligible. This approach is justified noticing that in the long-time limit of a quasi-stationary state, the proton current becomes small, as can be seen by computing the proton current given by Eq. (D7) for the case of uniform bacterial density using the exact solution Eq. (B12):

$$\bar{J}_{in} = \frac{Dc_0}{\sqrt{\pi Dt}}; \quad \text{for} \quad t \gg \tau^{2D} = \frac{D}{\alpha^2 h^2 \rho_0^2}. \tag{D12}$$

Thus we begin by positing that diverges as $v$ approaches zero, while $a \simeq a_0$ is approximately independent of $v$, and $q^2 \simeq q_0 v^\gamma$. By substituting these expressions into Eq. (D10a) and requiring that the equation holds in the asymptotic limit $v \to 0$, we obtain:

$$b = (1+a_0)^{\frac{\xi}{\xi-1}} \left(\frac{1}{v}\right)^{\frac{1}{\xi-1}}. \tag{D13}$$

Now, given that $a$ is much smaller than $b$ and significantly larger than $v$, Eq. (D10b) reduces to $4bk^2 = a$. Based on this simplified form, it follows that:

$$q^2 \simeq \frac{a}{4b} \simeq \frac{a_0}{4(1+a_0)^{\frac{\xi}{\xi-1}}} v^{\frac{1}{\xi-1}}. \tag{D14}$$

Finally, to determine $a_0$ we use Eq. (D10c) where we neglect $a$ compared to $b$, and substitute Eq. (D14) for $q^2$. The solution of the resulting equation is

$$a_0 \simeq \left(\frac{\xi \alpha N}{4\pi h D} - 1\right)^{-1}. \tag{D15}$$

Substituting Eqs. (D13) and (D14) in (D11) we arrive at:

$$\rho(r) = \rho_{max} \left[ 1 + \left(\frac{1+a_0}{v}\right)^{\frac{1}{\xi-1}} \tanh^2\left( \sqrt{\frac{\alpha \rho_{max}}{4D} \frac{a_0}{a_0+1}} \left(\frac{v}{1+a_0}\right)^{\frac{1}{\xi-1}} r \right) \right]^{-\xi}. \tag{D16}$$

To simplify this formula, it is useful to focus on the center of the condensate by expanding the hyperbolic tangent function to the leading order in $r$. In this approximation, the $v$ dependence of $\rho(r)$ cancels out, and (D16) simplifies to Eqs. (2.5) and (2.6).

The above calculation demonstrates that the main body of the bacterial condensate's distribution does not depend on the incoming flux; therefore, it is practically time-independent. The time-dependent aspects of the condensate distribution are confined to the far tails of the distribution, where they manifest themselves in a very gradual manner. This slow temporal change is described by $v^{\pm 1/(\xi-1)} \sim t^{\pm 1/(2\xi-2)}$ for a two-dimensional geometry. Thus, for $\xi = 6$, the time dependent component in the far tales of the distribution function is a function of $t^{\pm 1/10}$.

Formula (D16) is designed to provide an accurate approximation for bacterial density, both at the accumulation core and in regions farther away, where the density stabilizes to a near-constant value. While this approximation performs well for the channel geometry (as shown in the left panel of Fig. 8), it does not match the experimental data as closely in the case of disk-like geometry (right panel of Fig. 8). The reason for the better fit in stripe geometry, compared to disk geometry, is unclear. This discrepancy may stem from additional factors influencing the density profile, such as oxygen availability in densely populated regions. Supporting this hypothesis, Fig. 9 shows that the number of bacteria within condensates is considerably higher in the disk geometry compared to the channel geometry. Furthermore, when we repeat the experiment in a channel geometry with a higher initial cell count, the condensate shape resembles that observed in the disk-like geometry, as shown in Fig. 10. These findings suggest that the observed discrepancy is due to high-density effects not accounted for in our model, such as volume-filling constraints and aerotaxis. Thus, we conclude that the discrepancy is due to effects that emerge at high cell counts not accounted for by our model, such as volume filling and aerotaxis. Evidence that aerotaxis plays an important role in high-density condensates has been shown in Ref. [13].

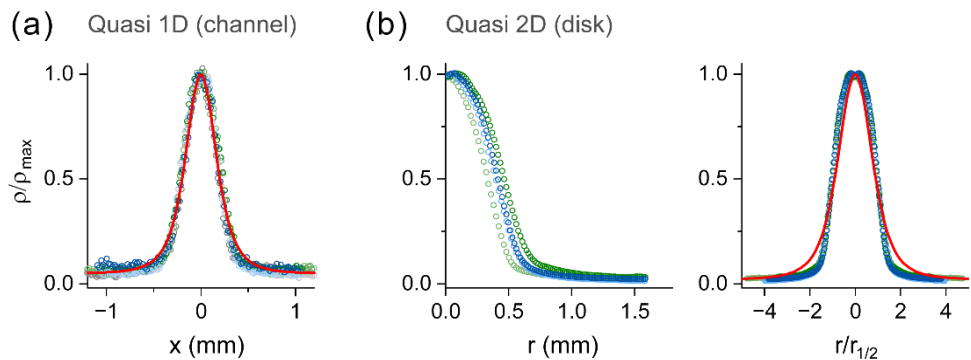

**Figure 8. Theory versus data for quasi-1D and quasi-2D geometries.** The density of condensates as a function of position for quasi-1D (a) and quasi-2D (b) geometries of the bacterial domain. The experimental setup use in (b) is shown in Fig. 7a. Data is represented by the colored dots and the solid red lines are the best fits to formula (D16). The densities are normalized by their maximal values. In the right panel of (b) the spatial coordinate was also normalized and plotted symmetrically for clarity. Evidently, experiments in the quasi-1D setup show little variability in the width of the condensates and aligns with the theory, while in the quasi-2D setup there are large fluctuations in the width but with a single shape function that deviates from the theory. The deviation is presumably due to the significantly elevated number of cells in each condensate (see discussion above and Fig. 9).

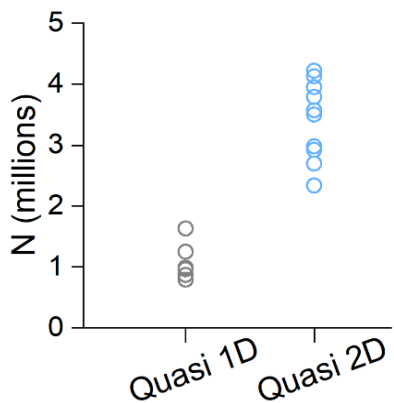

**Figure 9. The number of cells in condensates formed in different well geometries.** The total number of cells, $N$, in the quasi-1D setup (Fig. 2) and in the 2D disc setup is shown. The number of cells was calculated by summing the fluorescence intensity of each condensate, calibrated to units of $OD_{600}$. Under our experimental conditions, $OD_{600} = 0.5$ corresponds to approximately $7.2 \times 10^9$ cells/mL. Six condensates are shown in the quasi-1D setup (from three biological replicates, as in Fig. 2) and ten condensates are shown in the 2D setup. The number of cells in each condensate was significantly higher in the 2D setup compared to the quasi-1D setup.

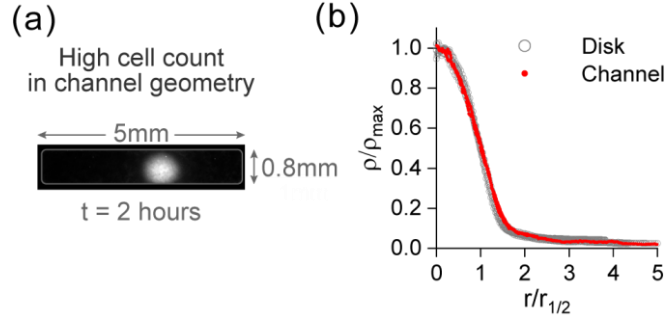

**Figure 10. Condensation in a channel geometry at high cell density.** Experiments were conducted in a channel geometry ($5\times0.8\times0.1\,\text{mm}^3$) with high initial cell density ($OD_{600}$ 1.2) to increase the total cell count ($N$) in the condensates, (a) Snapshot image of the bacterial layer at t = 2 hours. (b) Cell density profiles of condensates in these experiments (red lines) compared with the profiles in a disk geometry (gray markers). Profiles were normalized as in Fig. 8b. We estimate the total cell count in each condensate in the channel geometry to be $N \approx (3-4.5)\times10^6$ (see Fig. 8). At this elevated cell density, the condensates in the channel geometry exhibit shapes and sizes similar to those in the disk geometry, deviating from the model's predictions.

Before concluding this section, we aim to clarify the role of $K$ and demonstrate that it acts as a regularizer, preventing divergent solutions. To this end, it is instructive to calculate the distribution for the scenario where $K$ is set to zero. In that case, the solution of (D2) is $\rho = \mu c_{\mathcal{D}}^{-\xi}$, where $\mu$ is some constant. Focusing exclusively on the core body of the bacterial density, where the proton current flow, $J_{in}$, does not significantly influence the solution, we can now replace Eq. (D8) with:

$$\frac{1}{r}\frac{d}{dr}\left(r\frac{dc_{\mathcal{D}}(r)}{dr}\right) - \frac{\mu\alpha}{D}\left(\frac{1}{c_{\mathcal{D}}(r)}\right)^{\xi-1} = 0. \tag{D17}$$

Substituting a solution in the form $c_{\mathcal{D}}(r) = ar^{\sigma}$ yields two equations for the constants $a$ and $\sigma$ whose solution is

$$\sigma = \frac{2}{\xi} \quad \text{and} \quad a = \left(\frac{\mu\alpha\xi^2}{4D}\right)^{\frac{1}{\xi}}. \tag{D18}$$

Then substituting this solution in $\rho = \mu c_{\mathcal{D}}^{-\xi}$, one obtains a divergent solution:

$$\rho(r) = \frac{4d}{\alpha\xi^2 r^2}. \tag{D19}$$

A finite value of $K$ does not support a divergent solution. Thus, $K$ effectively functions as a regularizer, stabilizing the system by preventing unbounded behavior in the solutions.

**APPENDIX E: NUMERICAL SOLUTION**

The modified Keller-Segel equations [Eqs. (1.1)] were solved numerically assuming radial symmetry for both protons and bacteria. The equation for the dynamics of the protons (1.1a) was solved in three dimensions, while the equation for the dynamics of the bacteria (1.1b) was solved in two dimensions. Spatial derivatives were calculated with 2[nd] order accuracy and time integration was performed by using an explicit 4[th] order Runge-Kutta method. Proton uptake in the thin bacterial layer was accounted for by imposing absorbing boundary conditions for the protons:

$$\mathbf{J}_{in} \cdot \hat{\mathbf{z}}\big|_{z=0, r<d/2} = h\alpha\rho c, \tag{E1}$$

where $\mathbf{J}_{in}$ denotes the proton flux. We set the diameter of the two-dimensional bacterial layer to $d = 5\text{mm}$, and its thickness to $h = 120\mu\text{m}$. All other boundary conditions were set to ensure no cell or proton flux at their respective boundaries. The initial conditions were set such that bacteria and protons are uniformly distributed.

We ensured the accuracy of the numerical solution in two ways. First, we verified convergence by ensuring that the output remained unchanged when reducing the duration of time steps. Second, we compared the numerical solution for the proton distribution with the analytical one obtained for a homogeneous cell distribution (shown in Fig. 6). Comparison between the numerical solution for the shape of a mature bacterial condensate and the analytical approximation (formula D16) is plotted in Fig. 11.

The parameters used in the simulations are detailed in Table 1, and their detailed estimation was described previously (Table M1 in [13]). Notice that, for simplicity, the proton consumption, described by the parameter $\alpha$, was set to be independent of the proton concentration. Setting $\alpha \propto \sqrt{c}$ as in [13] did not have qualitative effects on the shape of condensates. Here, we set a constant $\alpha$ to match the magnitude of the condensation, while keeping the estimate bounded by the experiments in [13]. The chosen value of $\alpha$ resulted in longer time for the condensation to mature ($t_f$ = 15 hours).

| |
|---|
| $D = 5 \cdot 10^{-6}\ cm^2 s^{-1}$ |
| $\kappa = 30 \cdot 10^{-6}\ cm^2 s^{-1}$ |
| $K = 10^{-6}\ M\ (pH\ 6)$ |
| $D_c = 5 \cdot 10^{-6}\ cm^2 s^{-1}$ |
| $\alpha = 1.34 \cdot 10^{-3}\ s^{-1}/OD_{600}\ 0.4$ |

**Table 1.** Summary of the parameters used in the numerical simulations. $\alpha$ was chosen such that the magnitude of the condensation in the numerical solution matches that of the experiments. All other parameters were estimated using available experimental data (see Table M1 in [13] for details).

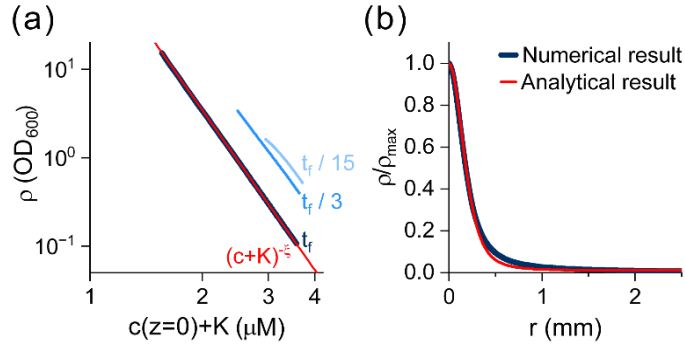

**Figure 11. Comparison of numerical solution with the analytical formula (D16).** (a) convergence to a power-law relation between the cell density and the proton distribution within the cell layer, $\rho \propto (c+K)^{-\xi}$. The relation is plotted at various times, shown as fractions of $t_f$, the time at which the shape of the bacterial condensate has saturated (here $t_f = 15$ hours). The red line is a plot of a line proportional to $(c+K)^{-\xi}$. Our numerical calculations confirm that the power law relation holds at the quasi-steady state, and evidently, even earlier. (b) Comparison between the numerical and the analytical solutions (formula D16). The analytical shape was fitted to the numerical solution by fitting $\rho_N = \left[1 + a\tanh^2(r/\ell)\right]^{-\xi}$ while enforcing $\xi = 6$ (as in the simulations). Note that the analytical solution matches the asymptotic behavior of the numerical solution.

**APPENDIX F: CONDENSATE MERGING DYNAMICS**

In this appendix, we derive Eqs. (2.7) and (2.8), which describe the dynamics of the merging of two bacterial condensates. However, before presenting the calculations, we first provide experimental evidence demonstrating that the observed merging dynamics result from genuine chemotactic attraction rather than an artifact of boundary effects. Specifically, we show that the motion of the condensates is not driven by attraction to a nearby boundary. To validate this, we conducted two additional experiments under conditions similar to those in Fig. 3 (initial conditions of $OD_{600}$ = 0.6 and pH 5.2) but with different confinement geometries to assess the role of boundary effects. The first geometry consisted of an isolated triangular well, while the second included two triangular wells connected by a 1 mm-long channel. In both cases, bacterial condensates formed at the predicted location (the center of the triangular well) within approximately 30 minutes. However, their subsequent movement was slow and lacked a preferred direction, as shown in Fig. 12. These results clearly demonstrate that condensate merging is not an artifact of the well geometry or driven by attraction to the boundary.

To analyze the merging dynamics of two bacterial condensates, we start with Eqs. (D3) and (D6), neglecting the incoming flux of protons. This approach is justified, as by the time the bacterial condensates form, this flux has already diminished significantly, and we are focused on much later stages of the process. Thus, our starting point is the equations:

$$\frac{\partial \rho}{\partial t} = \nabla_\perp \cdot \left[ D \nabla_\perp \rho + \kappa \rho \nabla_\perp \ln(K + c_\mathcal{D}) \right], \tag{F1}$$

$$D \nabla_\perp^2 c_\mathcal{D} - \alpha c_\mathcal{D} \rho = 0, \tag{F2}$$

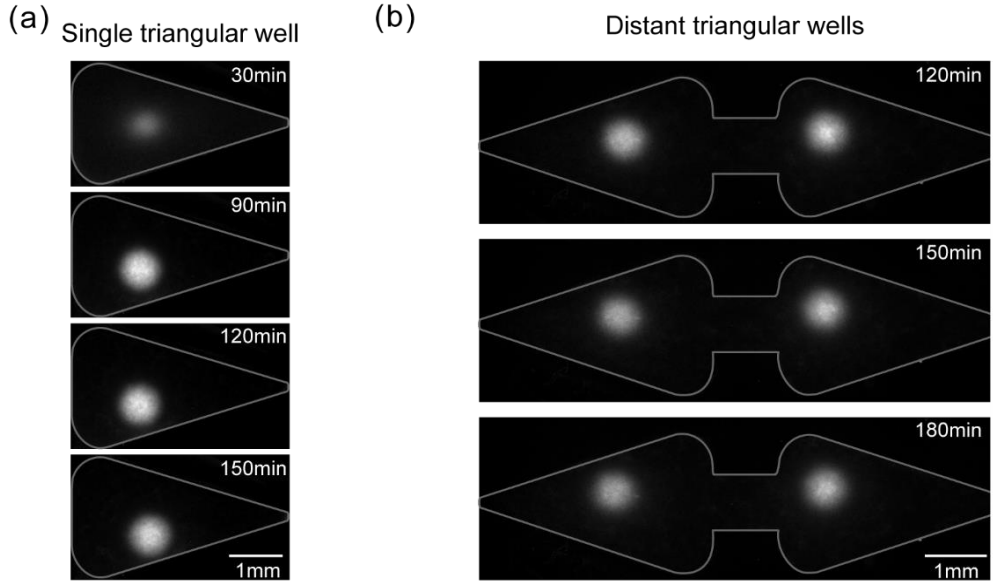

**Figure 12 | Coalescence of condensates is not due to the well geometry.** Bacterial clustering experimetes conducted under the same initial conditions as in Fig. 3, but with modified geometries. (a) Condensation in an isolated triangular well. Snapshots of the bacterial layer at different time points (as labeled) from a representative experiment. Condensates formed at the expected location ( at $t = 30\,\text{min}$ ) but displayed slow and undirected movement thereafter. (b) Condensation in distant, but connected, triangular wells. Two triangular wells were connected by a 1 mm-long channel. In these experiments, condensates formed approximately 3 mm apart and remained stationary, showing no movement toward each other.

where $\nabla_\perp$ represents the two-dimensional gradient. Here we retain only the time derivative of $\rho$, assuming that the proton rapidly adjusts to changes in the bacterial distribution. This approximation is justified by the observation that condensate merging is a collective phenomenon occurring over a time scale significantly longer than the typical diffusion time across a single condensate (as will be demonstrated below).

Representing the fields as $c_\mathcal{D} = \langle c \rangle + \delta c(\mathbf{r},t)$ and $\rho = \langle \rho \rangle + \delta\rho(\mathbf{r},t)$, where $\langle c \rangle$ and $\langle \rho \rangle$ are constants representing spatial averages (which remain constant over time), Eq. (F2) simplifies to:

$$D\nabla_\perp^2 \delta c = \alpha \left( \langle c \rangle + \delta c \right)\left( \langle \rho \rangle + \delta \rho \right) = \alpha \langle c \rangle \rho + \alpha \langle \rho \rangle \delta c . \qquad (F3)$$

Here, we have neglected the non-linear term $\delta c \delta \rho$ on the right-hand side, justified by the assumption that the merging dynamics are slow and changes in $\rho$ are gradual. As we will see, the precise value of the constant $\langle c \rangle$ has minor significance, whereas $\langle \rho \rangle$ plays an important role and will be determined in the following discussion. The formal solution to the equation above is:

$$\delta c = \frac{\alpha \langle c \rangle}{D\nabla_\perp^2 - \alpha \langle \rho \rangle} \rho . \qquad (F4)$$

Substituting this solution in Eq. (F1) we obtain the equation for the bacteria density

$$\frac{\partial \rho}{\partial t} = \nabla_\perp \cdot \left[ D\nabla_\perp \rho + \frac{\varpi \kappa \alpha}{D} \rho \nabla_\perp \frac{1}{\nabla_\perp^2 - \alpha \langle \rho \rangle/D} \rho \right]. \qquad (F5)$$

To obtain this result have used the following approximations:

$$\nabla_\perp \ln(K + c_\mathcal{D}) = \frac{\nabla_\perp c_\mathcal{D}}{K + \langle c \rangle + \delta c} \simeq \frac{\nabla_\perp \delta c}{K + \langle c \rangle} = \frac{\nabla_\perp}{K + \langle c \rangle} \frac{\alpha \langle c \rangle}{D\nabla_\perp^2 - \alpha \langle \rho \rangle} \rho$$
$$= \frac{\langle c \rangle}{K + \langle c \rangle} \nabla_\perp \frac{\alpha}{D\nabla_\perp^2 - \alpha \langle \rho \rangle} \rho \simeq \nabla_\perp \frac{\varpi \alpha}{D\nabla_\perp^2 - \alpha \langle \rho \rangle} \rho \qquad (F6)$$

Here, for the second approximation, we have neglected terms that are quadratic in $\delta c$. We also introduce the dimensionless parameter $\varpi = \langle c \rangle / (\langle c \rangle + K)$ which is positive and smaller than one.

The second term on the right-hand side of Equation (F5) represents the density-density interaction. If the neglected nonlinear terms were included, this term would dictate the steady-state shape of a single condensate. However, our current analysis focuses on the interaction between two distant colonies. In this context, it's important to note that the interaction term exhibits a screened density-density interaction, where $\lambda = \sqrt{D/(\alpha \langle \rho \rangle)}$ denotes the screening length. The rapid decay in the density of a single bacterial condensate allows us to disregard these nonlinear terms in our present discussion.

Eq. (F5) can also be manipulated and be written as a sum of two terms as follows:

$$\frac{\partial \rho}{\partial t} = \nabla_\perp \cdot \left\{ D\nabla_\perp \rho + \frac{\varpi \kappa \alpha}{D} \nabla_\perp \left[ \rho \frac{1}{\nabla_\perp^2 - \lambda^{-2}} \rho \right] \right\} - \frac{\varpi \kappa \alpha}{D} (\nabla_\perp^2 \rho) \frac{1}{\nabla_\perp^2 - \lambda^{-2}} \rho \qquad (F7)$$

To handle this equation, we express the operator acting on the density on the right-hand side of the equation in the form:

$$\frac{\varpi\kappa\alpha}{D}\frac{1}{\nabla_\perp^2 - \lambda^{-2}}\rho(\mathbf{r},t) = -\int d^2 r' G(\mathbf{r}-\mathbf{r}')\rho(\mathbf{r}',t), \qquad (F8)$$

where

$$G(\mathbf{r}-\mathbf{r}') \simeq \frac{\varpi\kappa\alpha}{2\pi D} K_0\left(|\mathbf{r}-\mathbf{r}'|/\lambda\right). \qquad (F9)$$

Here $K_0(z)$ is the zeroth order modified Bessel function of the second kind, also known as the MacDonald function.

To proceed, we consider two identical colonies positioned at a distance $2a(t)$ from each other, with the x-axis passing through the centers of these colonies. Assuming the bacterial density is symmetric with respect to the line $x = 0$, it is represented by:

$$\rho(\mathbf{r},t) = \rho_1[-\mathbf{r}-\mathbf{a}(t)] + \rho_1[\mathbf{r}-\mathbf{a}(t)], \qquad (F10)$$

where $\rho_1(\mathbf{r})$ denotes the density distribution of a single condensate (see also Fig. 13). Hereafter $\rho_1(\mathbf{r})$ will indicate the density of a single condensate, excluding the baseline bacterial density. This adjustment is made because the baseline density does not influence the dynamics of condensate merging. Consequently, the approximate density for a single condensate, $\rho_1(r)$, with the background subtracted, is given by Eq. (2.5) with $\ell$ given by Eq. (2.6). The screening length $\lambda$, up to a constant of order unity, is the width of a single condensate, $\ell$, which is the natural static length scale in the problem.

When the bacterial accumulations are far apart, their individual shape is approximately described by Eq.(2.5). However, as the condensates approach each other, their shapes may evolve over time. Despite this change, for the purpose of determining the function $a(t)$, detailed knowledge of the shape evolution is not necessary. The essential assumptions are that the center of mass of each condensate remains well-defined and that the number of cells within each condensate, $N$, remains constant, thus

$$N = h\int d^2 r \rho_1(\mathbf{r}), \quad \int d^2 r \mathbf{r} \rho_1(\mathbf{r}) = 0, \quad \text{and} \quad \int d^2 r\, r^2 \rho_1(\mathbf{r}) < \infty, \qquad (F11)$$

Next, we multiply Eq. (F7) by $x$ and integrate over half-space, $x > 0$

$$\int_{x>0} d^2 r x \frac{\partial \rho}{\partial t} = \int_{x>0} d^2 r x \nabla_\perp \cdot \left\{ D\nabla_\perp \rho + \frac{\varpi\kappa\alpha}{D}\nabla_\perp\left[\rho\frac{1}{\nabla_\perp^2 - \ell^{-2}}\rho\right]\right\}$$
$$+ \frac{\varpi\kappa\alpha}{D}\int_{x>0} d^2 r x (\nabla_\perp^2 \rho)\frac{1}{\nabla_\perp^2 - \ell^{-2}}\rho . \qquad (F12)$$

When substituting expression (F10) into the above integrals, it becomes clear that the interaction terms consist of two types: The first is self-interaction within a single condensate, which, together with the diffusive term $D\nabla_\perp^2 \rho$ and the neglected nonlinear terms, determines the shape of a single condensate. The second type comprises interaction terms that involve the two distinct condensates. For the purpose of calculating the attraction between the accumulations, only these latter terms should be considered. Thus, the relevant part of the integral that addresses the inter-condensate interactions is:

$$\int d^2rx \frac{\partial \rho_1[\mathbf{r}-\mathbf{a}(t)]}{\partial t} = \frac{\varpi\kappa\alpha}{2\pi D}\int d^2rx \nabla_\perp \cdot \left\{ \nabla_\perp \left[ \rho_1[\mathbf{r}-\mathbf{a}(t)] \int d^2r' K_0(|\mathbf{r}-\mathbf{r}'|/\ell)\rho_1[-\mathbf{r}'-\mathbf{a}(t)] \right] \right\}$$
$$+ \frac{\varpi\kappa\alpha}{2\pi D}\int d^2rx \left( \nabla_\perp^2 \rho_1[\mathbf{r}-\mathbf{a}(t)] \right) \int d^2r' K_0(|\mathbf{r}-\mathbf{r}'|/\ell)\rho_1[-\mathbf{r}'-\mathbf{a}(t)] \quad (F13)$$

Here, we extend the integration region to encompass the entire space, as $\rho_1(\mathbf{r})$ is localized and we assume $2a(t) > \ell$. This approximation is also directly justified by evaluating the integral in (F12) and demonstrating that the terms excluded in (F13) are small. This smallness arises either because they are proportional to $N$ rather than $N^2$ or because they are smaller by an additional factor of $(\ell/a)^{2\xi-1}$ compared to the results of the second integral on the right-hand side of Eq, (F13).

The left-hand side of Eq. (F13) yields

$$\int d^2rx \frac{\partial \rho_1[\mathbf{r}-\mathbf{a}(t)]}{\partial t} = \frac{N}{h}\frac{da(t)}{dt}. \quad (F14)$$

The first term on the right-hand side of (F13) can be evaluated using integration by parts. The surface term vanishes, and the remaining integrand, which is a total derivative, is zero because the bacteria density decays to zero far from the condensate center.

To evaluate the second term on the right-hand side of (F13), we take into account that $\rho_1(\mathbf{r})$ is a localized function. Therefore, the integral over $\mathbf{r}'$ can be approximated by the following expression:

$$\int d^2r' K_0\left(\frac{|\mathbf{r}-\mathbf{r}'|}{\ell}\right)\rho_1[-\mathbf{r}'-\mathbf{a}(t)] \simeq \frac{N}{h} K_0\left(\frac{|\mathbf{r}+\mathbf{a}(t)|}{\ell}\right). \quad (F15)$$

A similar approach for the integral over $\mathbf{r}$ is not effective because the integrand becomes a total derivative after one integration by parts. Therefore, one has to expand $K_0(|\mathbf{r}+\mathbf{a}|/\ell)$ to linear order around $\mathbf{r}=\mathbf{a}$ to manage the calculations effectively. Thus,

$$K_0\left(\frac{|\mathbf{r}+\mathbf{a}|}{\ell}\right) \simeq K_0\left(\frac{2|\mathbf{a}|}{\ell}\right) - K_1\left(\frac{2|\mathbf{a}|}{\ell}\right)\frac{\mathbf{a}\cdot\mathbf{r}}{2\ell a} = K_0\left(\frac{2a}{\ell}\right) - K_1\left(\frac{2a}{\ell}\right)\frac{x}{\ell}. \quad (F16)$$

Substituting this expansion in the second integral on the right-hand side of Eq. (F13), and evaluating the integral by parts we obtain

$$\frac{\varpi\kappa\alpha}{2\pi D}\int d^2rx \left(\nabla_\perp^2 \rho_1[\mathbf{r}-\mathbf{a}(t)]\right)\int d^2r' K_0\left(\frac{|\mathbf{r}-\mathbf{r}'|}{\ell}\right)\rho_1[-\mathbf{r}'-\mathbf{a}(t)]$$
$$\simeq \frac{\varpi\kappa\alpha}{2\pi D}\int d^2rx \nabla_\perp^2 \rho_1(\mathbf{r}-\mathbf{a}) N_1 K_0\left(\frac{|\mathbf{r}+\mathbf{a}|}{\ell}\right) \quad (F17)$$
$$\simeq \frac{N}{h}\frac{\varpi\kappa\alpha}{2\pi D}\int d^2rx \nabla_\perp^2 \rho_1(\mathbf{r})\left[ K_0\left(\frac{2a}{\ell}\right) - K_1\left(\frac{2a}{\ell}\right)\frac{x}{\ell}\right] = -\frac{\varpi\kappa\alpha N^2}{\pi\ell D h^2} K_1\left(\frac{2a}{\ell}\right)$$

Combining (F14) and (F17), we arrive the equation for the separation between the accumulations:

$$\frac{d2a}{dt} = -\frac{2N\varpi\kappa\alpha}{\pi\ell hD} K_1\left(\frac{2a}{\ell}\right) = -\frac{\ell}{\tau_{merge}} K_1\left(\frac{2a}{\ell}\right), \tag{F18}$$

where

$$\frac{1}{\tau_{merge}} = \frac{2N\varpi\kappa\alpha}{\pi h\ell^2 D} = 8\pi\varpi\frac{D}{\ell^2}. \tag{F19}$$

To obtain this result we used estimation (D5a) for $N$, and the definition $\xi = \kappa/D$ (see Eq. (2.5)).

The asymptotic behavior of the modified Bessel function at large arguments, $a > \ell$, where the above derivation holds, is given by $K_1(2a/\ell) \simeq \sqrt{\pi\ell/4a}\exp(-2a/\ell)$. Up to logarithmic accuracy, we may neglect the prefactor, and thus Eq. (F18) simplifies to:

$$\exp\left(\frac{2a}{\ell}\right)\frac{d2a}{dt} = -\frac{\ell}{\tau_{merge}}. \tag{F20}$$

The solution to this equation is given by Eq. (2.8).

Finally, notice that although the above calculation does not assume that bacterial accumulations retain their shape during merging, our experimental data, shown in Fig. 13, suggest that this is approximately the case up until the point of merging.

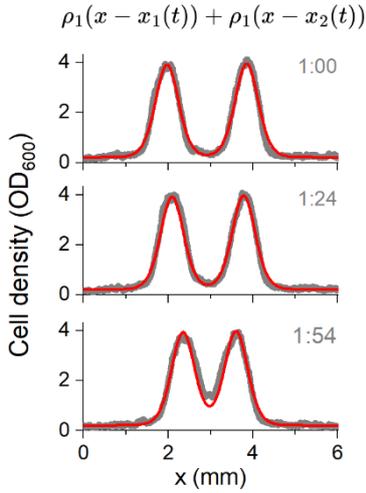

**Figure 13. Condensates maintain their shape during coalescence.** The density profiles from the experiment in Fig. 3a-c is shown in gray at various time points (as labeled in hours). The red lines describe $\rho_1[x-x_1(t)] + \rho_1[x-x_2(t)]$ where $\rho_1$ is the shape of a single condensate, given by Eq. (2.5), while $x_1$ and $x_2$ are the centers of the corresponding condensate. Note that in the fitted red lines $\rho_1$ was the same for both condensates and only the values of $x_1$ and $x_2$ were changed to fit different time points. Despite the tails of the condensates already in contact, their shape remained nearly constant during coalescence.

## APPENDIX G: IMPACT OF THE DEPENDENCE OF THE CONSUMPTION COEFFICIENT $\alpha$ ON PROTON DENSITY ON SYSTEM BEHAVIOR

The assumption that the weak dependence of the parameter $\alpha$ on proton density can be neglected was necessary to derive the exact solution describing the evolution of the proton density profile in the direction perpendicular to the bacterial domain, where the lateral spatial chemotactic instability develops (Appendix B). Numerical solutions of Eqs. (2.1) with $\alpha = \alpha_0\sqrt{c}$ that demonstrate: (a) The system's qualitative behavior is largely unaffected by this dependence, and (b) With the known value of $\alpha_0$ (determined in [13]), the numerical results align well with experimental findings – both in the time scales, magnitude and shape of the condensation - without the need for fitting parameters [13] (see Fig. 14). Evidently, the inclusion of the corrected consumption term, $\alpha = \alpha_0\sqrt{c}$, did not have a qualitative affect on the shape of condensates, and does not explain the discrepancy between the theory and experiments at high cell count (Figs. 8-10 in Appendix D). Note that in the experiments, we see a shallow baseline of fluorescent cells (about ~5% of the peak height). These non-cooperating cells, presumably non-motile, are not considered in the simulations, and thus, the baseline decays to a different value than in the experiments. However, both the experiments and simulations can be fitted by formula D16 (data-theory fits are shown in Figs. 2 and 8; theory-simulations fits are shown in Figs. 11 and 14).

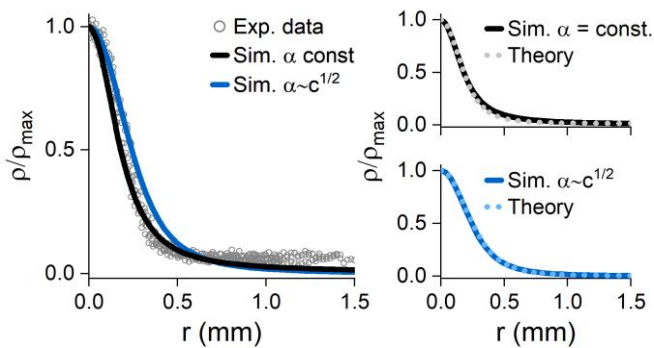

Figure 14. Impact of the consumption term on the shape of mature condensates. Equations (1.1) were solved using a constant $\alpha$ (as in Appendix F), shown as black lines or $\alpha = \alpha_0 \sqrt{c}$ (taken from [13]), shown as blue lines. Left – comparison between the simulations and the experimental data. Right – fits between the simulations and the theoretical shape (formula D16).